\documentclass[12pt,a4paper]{article}
\usepackage{float}
\usepackage[english]{babel}
\usepackage{latexsym}
\usepackage {amssymb}
\usepackage {amsmath}
\usepackage {mathtools}
\usepackage{color}
\usepackage{hyperref}
\usepackage{setspace}
\usepackage{booktabs}
\usepackage[]{algorithm2e}
\usepackage{multirow}
\usepackage[titletoc,title]{appendix}

\usepackage{rotating}

\usepackage[numbers]{natbib}
\bibpunct{(}{)}{;}{n}{,}{,}
\renewcommand{\cite}{\citep}

 \DeclareMathOperator{\diag}{diag}

\def\beqn{\begin{eqnarray*}}
\def\eeqn{\end{eqnarray*}}
\def\beq{\begin{eqnarray}}
\def\eeq{\end{eqnarray}}
\def\bm#1{\mbox{\boldmath{$#1$}}}

\def\diag{\mbox{diag}}

\newcommand{\im}{\bm{i}}

\newtheorem{lemma}{Lemma}

\makeatother
\makeatletter
\renewcommand{\@fnsymbol}[1]{\@arabic{#1}}
\makeatother
\title{Joint Non-parametric Estimation of Mean and Auto-Covariances for Gaussian Processes}

\author{
Tatyana Krivobokova
\footnote{
Department of Statistics and Operations Research, 
University of Vienna, 
Austria.}
\and
Paulo Serra
\footnote{
Department of Mathematics,
Vrije Universiteit Amsterdam, 
the Netherlands.}
\and
Francisco Rosales
\footnote{
Graduate school of Business,
Universidad ESAN,
Peru.}
\and
Karolina Klockmann
\footnote{
Department of Statistics and Operations Research, 
University of Vienna, 
Austria.}
}

\begin{document}

\baselineskip=25pt

\maketitle

\begin{abstract}
\baselineskip=15pt
\noindent
Gaussian processes that can be decomposed into a smooth mean function and a stationary autocorrelated noise process are considered and a fully automatic nonparametric method to simultaneous estimation of mean and auto-covariance functions of such processes is developed. Our empirical Bayes approach is data-driven, numerically efficient and allows for the construction of confidence sets for the mean function. 
Performance is demonstrated in simulations and real data analysis. 
The method is implemented in the R package {\sl eBsc} that accompanies the paper.\\

\end{abstract}

\newpage

\section{Introduction}\label{sec:introduction}


Consider the following observations from a fixed design, nonparametric regression model
\beq
\begin{aligned}
&&Y_i=f(t_i)+\sigma \epsilon_i,\;\mathbb{E}(\epsilon_i)=0,\;\sigma>0,\;\;i=1,\ldots,n\\
&&\mbox{corr}(\epsilon_i,\epsilon_j)=r(i-j)=r_{|i-j|}\in[-1,1],\;\;i,j=1,\ldots,n,
\label{eq:model}
\end{aligned}
\eeq
where $r(\cdot)$ denotes the autocorrelation function of the underlying noise process, and $r_{|i|}$ denotes autocorrelation at lag $i$.
The design points $t_i\in\mathbb{R}$ are equidistant and represent time points. The functions $f$ and $r$ are unknown, but we assume that the noise terms $\{\epsilon_i\}_{i=1}^n$ are sampled from a stationary noise process and that $f$ is a smooth function. 
(In this paper we use the Gaussian process as a template for such data, but the methodology can be applied to other processes as well.)
The observations $\{Y_i\}_{i=1}^n$ might be measures of some experimental quantity observed with a time dependent measurement error; in this case estimation of $f$ is of interest, while $r$ is considered as a nuisance parameter. It might also be that $\{Y_i\}_{i=1}^n$ are measurements from a stochastic process indexed by time or space with a seasonal or other deterministic effect, described by $f$; in this case the focus is rather on estimation of the autocorrelation $r$.


Non-parametric estimation of the covariance structure of a vector of observations with \emph{known} mean is already a challenging problem.
This problem is important in time series analysis -- e.g., for prediction -- as well as in multivariate analysis, in problems such as clustering, principal component analysis (PCA), linear- and quadratic discriminant analysis and regression analysis.
Two frameworks are usually considered when the mean is known: either
there are $n$ independent observations of a $p$-dimensional vector with correlated components, cf.~\cite{Bickel:2008a,Bickel:2008b}; or
there is one observation of an $n$-dimensional vector sampled from a stationary process, cf.~\cite{Xiao:2012}.
The second framework is the one relevant for our setting.
Irrespectively of the framework, natural (moment or sample auto-covariances) estimators for the covariance matrix of the observed vectors are well known to be inconsistent (in, e.g., operator-norm) and some form of regularisation is needed (e.g., banding, tapering, thresholding) to ensure positive definiteness and consistency of the estimator. To the best of our knowledge, there is no simple data-driven approach to select the regularisation parameter in this context, and thus fully taking the correlation structure into account; see, e.g., \cite{Pourahmadi:2011}.
Minimax rates of estimators of the correlation structure depend on the decay of the autocorrelation function as the lag increases (or, alternatively, smoothness of the corresponding spectral density); cf.~\cite{yang2001nonparametric}, \cite{Cai:2010}, \cite{Pourahmadi:2011}, \cite{Xiao:2012}, see also~\cite{fan2016overview}.
As such, non-parametric estimation of the covariance structure of the process is of great importance.

If the mean function is unknown, then a natural approach to covariance estimation is to obtain a consistent estimator for the mean function first and then apply methods for (nonparametric) covariance estimation with a known mean. Unfortunately, ignoring dependence on the error process hinders the estimation of $f$.

\begin{figure}[h!]
\begin{tabular}{cc}
\includegraphics[width=0.45\textwidth]{"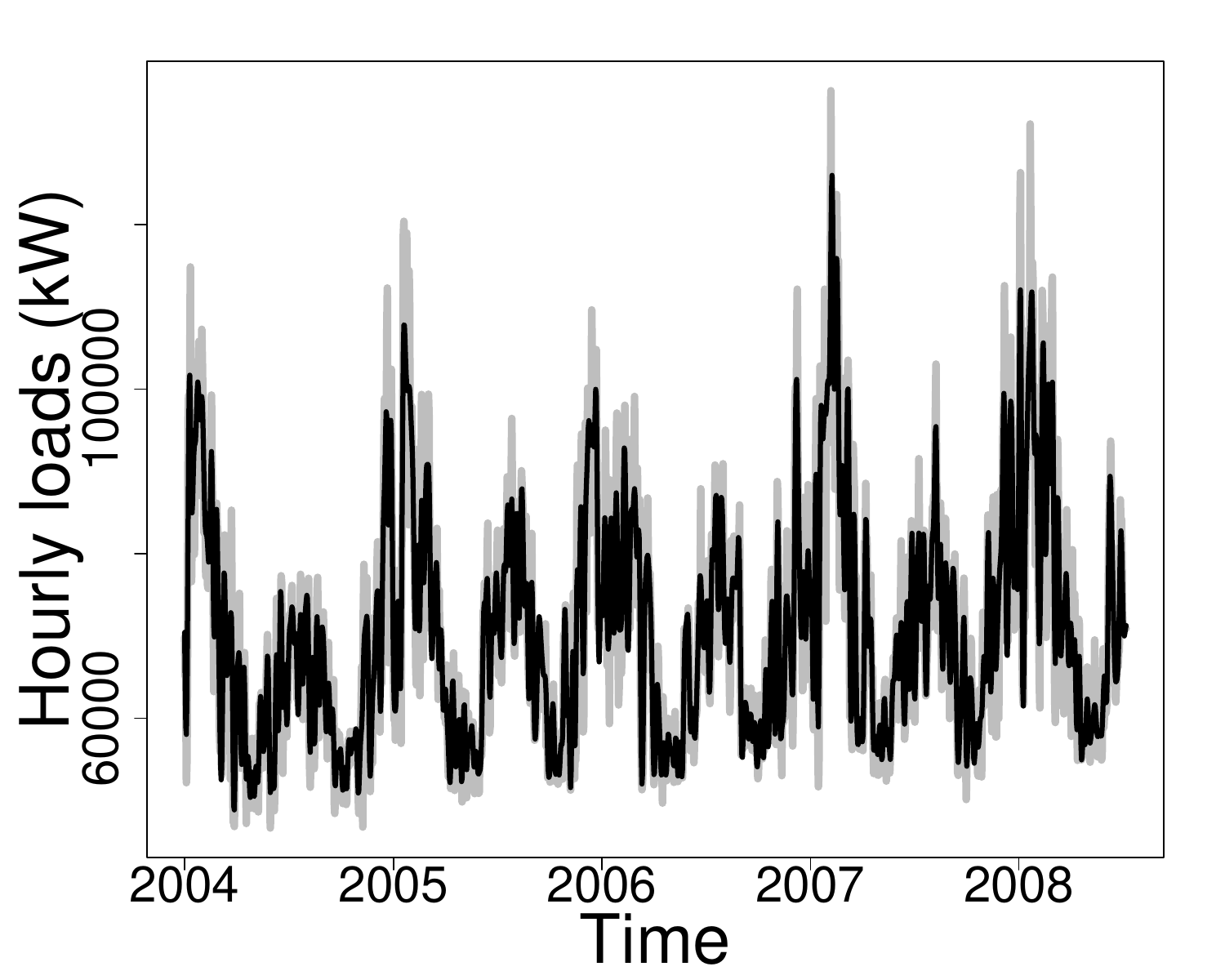"}&
\includegraphics[width=0.45\textwidth]{"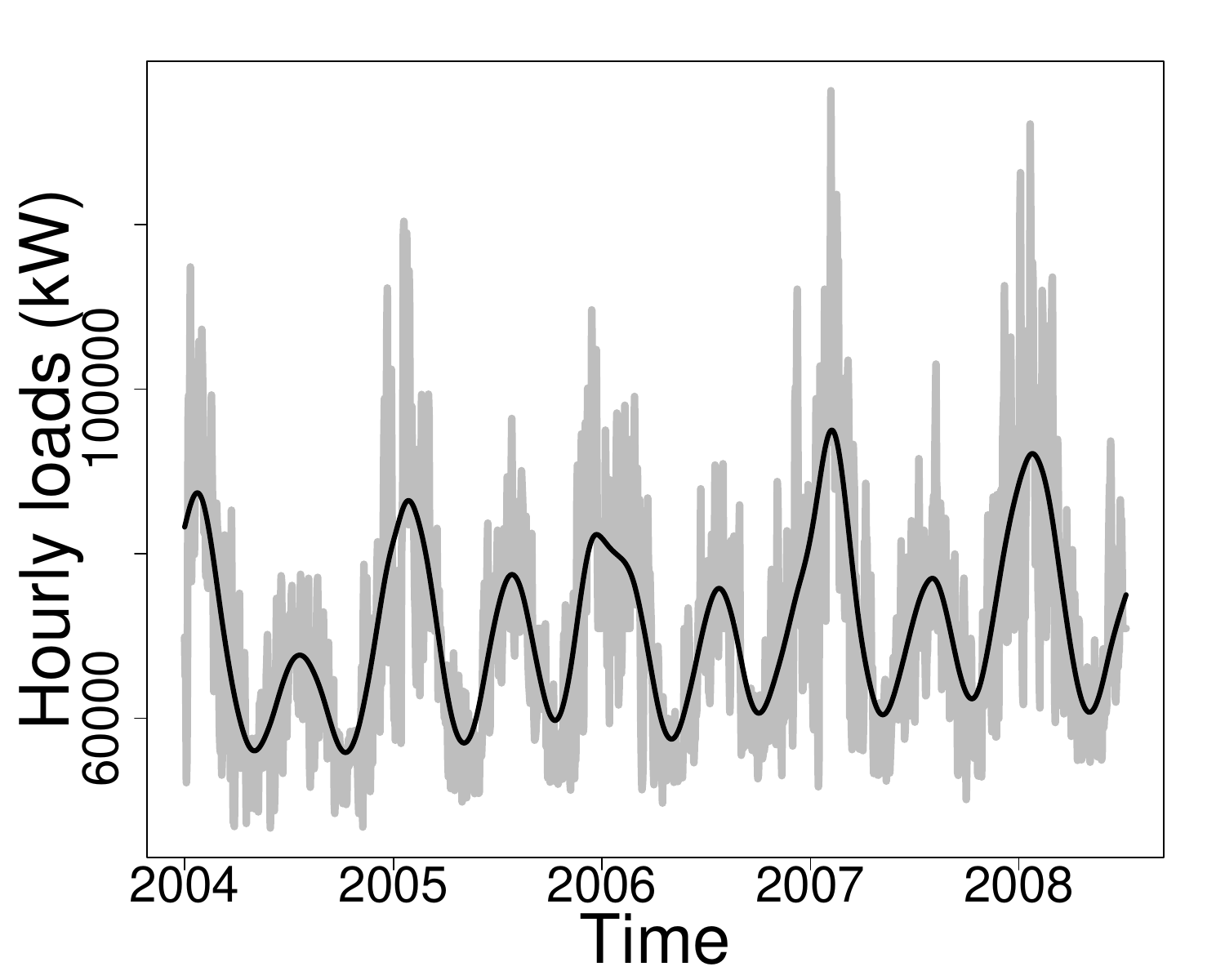"}\\
\end{tabular}
\caption{\small Hourly loads (kW) for a US utility estimated ignoring dependence in the data (left) and assuming that errors follow an autoregressive process of order one (right). Grey line shows the data and black lines show estimators. Estimators are obtained by functions {\tt gam} (left) and {\tt gamm} (right) of the R package {\sl mgcv}.}
\label{fig:intro}
\end{figure}

To exemplify the problem, consider a time series of hourly loads (kW) for a US utility (grey line in Figure \ref{fig:intro}), described in detail in Section \ref{sec:examples}. This process has clear seasonal effects over years and possibly over weeks and days. These deterministic effects can be modelled by a function $f$. If no parametric assumptions on $f$ is made and errors are treated as i.i.d., then all standard nonparametric estimators of $f$ are heavily affected by the ignored dependence in the errors. In particular, automatic selectors of the smoothing parameter (e.g., by cross-validation) choose a biased smoothing parameter leading to a nearly interpolating estimator in case of positively correlated $\{\epsilon_i\}_{i=1}^n$; see the black line on the left plot in Figure \ref{fig:intro}. This problem has been known for several decades; for an overview see \cite{Opsomer:2001}.

Hence, there is a need for an approach that allows for correct smoothing of the mean over a large family of possible auto-covariance functions.
Available methods rely on an explicit, parametric assumption on the correlation structure, e.g., assuming that $\{\epsilon_i\}_{i=1}^n$ follows an ARMA$(p,q)$ process. Once $r$ is parametrised, the usual smoothing parameter selection criteria are adjusted to incorporate $r$ and both $f$ and $r$ are estimated simultaneously. For example, \cite{Hart:1994} introduces a time series cross-validation for $f$ estimated by kernel estimators assuming errors to follow an AR(p) process, see also \cite{Altman:1990} and \cite{hall2003using}.  \cite{Kohn:1992} use spline smoothing for estimation of $f$ and a general ARMA$(p,q)$ model for the errors, estimating all parameters from either generalised cross validation or maximum likelihood. Similar ideas are employed in \cite{Wang:1998}.
This approach does allow for simultaneous estimation of $f$ and $r$, but is limited to parametric models for $r$. Of course, the true correlation structure might be much more complex than any parametric one and if the parametric model for the error process is misspecified, then the estimator of $f$ is strongly affected.
More importantly, in practice it is difficult to select and verify a parametric model for $r$, if the mean function $f$ is unknown.

Smoothing with low-rank splines and an ARMA$(p,q)$ model for the residuals is implemented in the  {\tt gamm} function of the R package {\sl mgcv}. This function has been used to estimate hourly loads in the right plot of Figure \ref{fig:intro}, assuming that the errors follow an AR(1) process. However, it is difficult to verify if this model is appropriate for the data at hand.
For more detailed analysis of this dataset see Section~\ref{sec:examples}.

Another group of methods for estimating $f$ is based on trying to eliminate the influence correlation has on the smoothing parameter thus treating $r$ as a nuisance parameter. For instance, \cite{Chu:1991} and \cite{Hall:1995} study the modified cross-validation criterion obtained by leaving out a whole block of $2l+1$ observations around each observation. \cite{Chiu:1989} and \cite{Hurvich:1990} study Mallow's $C_p$ and cross validation in the frequency domain. A different route is taken by \cite{Herrmann:1992} and \cite{lee2010bandwidth}: in the smoothing parameter selection criteria they incorporate some sample estimators of auto-covariances that depend on unknown parameters linked to assumptions on the error process. These methods require careful selection of tuning parameters, for which no data-driven methods are available.

In this paper we develop a likelihood based method that provides an estimator for a regression function $f$ observed with correlated, additive noise.
We also develop estimators of the noise level $\sigma$ and of the autocorrelation function $r$ as well.
Furthermore, our empirical Bayesian framework provides a computationally attractive way of constructing confidence sets for the regression function that take the correlation structure into account.
There are quite a few novel points in our work.
Contrary to other approaches in the literature, our method is completely automatic so that no tuning parameters need to be set by the user, and it is also fully nonparametric.
To the best of our knowledge, our estimate of the autocorrelations is also novel --
we use spline smoothers to estimate the spectral density of the noise process.
The autocorrelations are then reconstructed from the estimate of the spectral density, rather than by tapering or thresholding some empirical estimate. The resulting covariance matrix is positive definite by construction, and consistent in operator norm.
Further properties of this estimator require new tools to be developed and will be investigated in a separate work.

This paper is structured as follows.
In 
Section~\ref{sec:estimators} we introduce our estimators, 
Section~\ref{sec:numerical} contains simulation results, 
Section~\ref{sec:examples} presents a real data example, and 
Section~\ref{sec:conclusions} closes the paper with some conclusions.
Technical results are in the Appendix to this paper.

\section{Construction of the Estimators}\label{sec:estimators}

Assume that in model \eqref{eq:model} the regression function $f$ belongs to a Sobolev space $ \mathcal{W}_{\beta}$, $\beta>1/2$; see~\ref{app:DRB} for more details.
The noise terms $ \epsilon_i$ are sampled from a stationary, Gaussian noise process with zero mean and variance $\sigma^2>0$. 
The Gaussian model is used here as a template to construct estimators, but we work within a penalisation framework (cf.~\ref{eq:penalized_least_squares_criterium}) so that we expect our estimators to perform well under other data distributions.
To estimate $f$ we use the so called infill asymptotics, cf.~\cite{Robinson1989} for more details. 
Define $\bm{Y}=(Y_1,\dots,Y_n)^T$, $\bm{f}=f(\bm{t})=\{f(t_1), \dots,f(t_n)\}^T$, $\bm{\epsilon}=(\epsilon_1,\ldots,\epsilon_n)^T$, where $t_i=(i-1)/(n-1)$, $i=1,\ldots,n$.
We model the observations in~\eqref{eq:model} as
$$
\bm{Y}=\bm{f}+\sigma \bm{\epsilon},\;\;\;\bm{\epsilon}\sim N(\bm{0},\bm{R}).
$$
The symmetric $n\times n$ matrix $\bm{ R}$ satisfies the regularity assumptions described in Lemma~\ref{lemma:diag2}.
These do imply that the first row of $\bm{ R}$ is absolutely summable which means that we model the noise process as having short-term dependence.
However, the assumptions are otherwise quite mild; in particular, we consider a large (non-parametric) collection of covariance structures for the noise terms covering, for instance ARMA noise models as a very particular case.
We also assume that for some $0<\delta<1$, the eigenvalues of $\bm{R}$ lie on the interval $[\delta,\,1/\delta]$.



We estimate $f$ using a smoothing spline, i.e., we find $\hat{f}$ that solves
\begin{equation}\label{eq:penalized_least_squares_criterium}
\min_{f\in\mathcal{W}_q} \left[ \frac1n \big\{ \bm{Y} - f(\bm{t}) \big\}^T\bm R^{-1}\big\{ \bm{Y} - f(\bm{t}) \big\} + \lambda \int_0^1\big\{f^{(q)}(t)\big\}^2\,dt \right],
\end{equation}
for some $q\in\mathbb{N}$, $\lambda>0$ and some ``working'' correlation matrix $\bm R$. 
Subsequently, $q$, $\lambda$, and $\bm{R}$ are estimated from the data using the empirical Bayes approach.
It is well-known that for given $\bm R$, $q$ and $\lambda$, the resulting estimator $\hat{f}$ is a natural spline of degree $2q-1$ with knots at $\bm{t}$ and can be written as $\hat{{f}}(t)=\bm{S}(t)\bm{Y}$, where $\bm{S}$ is a $n\times n$ smoother matrix. To represent $\bm{S}$ we choose the so-called Demmler-Reinsch basis of the natural spline space of degree $2q-1$, which is defined in Section \ref{app:DRB} in the Appendix.
As such, we can write the smoother matrix $\bm S$ as
\beq\label{eq:SR}
\bm{S} =
\bm{S}_{\lambda,q,\bm R} &= \bm\Phi_q\left\{\bm\Phi_q^T\bm{R}^{-1}\bm\Phi_q+\lambda\diag(n{\bm{\eta}}_q)\right\}^{-1}\bm\Phi_q^T\bm{R}^{-1},
\eeq
where $\bm\Phi_q$ is an appropriate $n\times n$ basis matrix and $\bm\eta_q\in\mathbb{R}^n$ is a vector of eigenvalues.
This representation makes the dependence of $\bm{S}$ on the parameters $\lambda$, $\bm R$, and $q$ more explicit.
To keep the notation simple we omit the dependence on these parameters, unless these are set to a particular value.

\subsection{Bayesian Interpretation}\label{sec:estimators:bayes}

The estimator $\hat f$ has a Bayesian interpretation which provides us with a convenient way of estimating all of the unknown parameters by employing the empirical Bayes approach.

We start by endowing $\bm{f}$ with a partly informative prior --
given $(\bm{t},\lambda,q,\sigma^2, \bm{R})$, the prior on $\bm{f}$ admits a density proportional to
\beq\label{eq:prior_f}
\left|\frac{\bm{R}^{-1}(\bm{S}^{-1}-\bm{I}_n)}{2 \pi \sigma^2}\right|^{1/2}_+\exp\left\{ -\frac{\bm{f}^T\bm{R}^{-1}(\bm{S}^{-1}-\bm{I}_n)\bm{f}}{2\sigma^2} \right\},
\eeq
where $|\cdot|_+$ denotes the product of non-zero eigenvalues ($\bm{S}$ has exactly $q$ eigenvalues equal to $1$). This prior corresponds to a non-informative part on the null-space of $\bm{R}^{-1}(\bm{S}^{-1}-\bm{I}_n)$ and a proper Gaussian prior on the remaining space. (This prior distribution happens to be independent of $\bm{R}$. This follows from the identity  $\bm{R}^{-1}(\bm{S}^{-1}-\bm{I}_n)=\bm{S}_I^{-1}-\bm{I}_n$, where $\bm{S}_I$ denotes the smoother matrix with $\bm{R}=\bm{I}_n$; cf.~\ref{app:aux:matrix_idents} for the derivation of the identity.)
The mean of the corresponding posterior distribution for $\bm{f}|(\bm t,\lambda,\sigma^2,\bm R)$ is the smoothing spline estimator $\hat{\bm f\,\,}=\bm{S}\bm{Y}$; 
cf.~\cite{Speckman:2003}.
The variance $\sigma^2$ given $(\bm{t}, \lambda,q, \bm{R})$ is endowed with an inverse-gamma prior $\mbox{IG}(a,b)$, $a,b>0$.

The resulting prior on $(\bm f, \sigma^2)|(\lambda,q,\bm R)$ is conjugate for model~\eqref{eq:model} in the sense that the posterior distribution on $(\bm f, \sigma^2)|(\lambda,q,\bm R)$ is a known distribution.
Namely, the marginal posterior for $\sigma^2$ given $(\lambda, q, \bm R)$ is an inverse gamma distribution with shape parameter $(n-q+2a)/2$ and scale parameter $\{\bm Y^T \bm R^{-1}({\bm I_n} - \bm S)\bm Y+2b\}/2$. As for $\bm f|(\lambda, q, \bm R)$, its posterior distribution is a multivariate t-distribution with $n+1$ degrees of freedom, mean $\hat{\bm f\,\,}=\bm S \bm Y$, and scale $\hat\sigma^2 \bm S\bm R$.

It remains to estimate $\lambda$, $q$ and $\bm{R}$ which are parameters of the prior. To do so, we employ the empirical Bayes approach: we estimate these parameter from the marginal distribution of $\bm{Y}$, given $(\bm{t},\lambda,q, \bm{R})$, which is a multivariate $t$-distribution (cf.~\cite{Kotz:2004} for the definition of this distribution) with density
\[
\frac{\Gamma\{a+\frac{n-q}2\}\left|\bm{R}^{-1}(\bm{I}_n-\bm{S})\frac{2a-q}{2b}\right|_+^{1/2}}{\{\pi(2a-q)\}^{n/2}\Gamma(a-q/2)}\left\{1+\frac{\bm{Y}^T\bm{R}^{-1}(\bm{I}_n-\bm{S})\bm{Y}}{2b}\right\}^{-(n+2a-q)/2}.
\]
It remains to set the parameters $a$ and $b$. The choice of parameters $a$ and $b$ is irrelevant for the asymptotics (as long as $a$ and $b$ are $o(n)$ and lead to a proper prior and marginal distributions), so to simplify the log-likelihood, we set $b=1/2$ and $a=(q+1)/2$ to obtain
\begin{eqnarray}\label{eq:RES}
\ell_n(\lambda,q,\bm{R})=
-\frac{n+1}{2}\log\left\{\bm{Y}^T\bm{R}^{-1}(\bm{I}_n-\bm{S})\bm{Y}+1\right\}+\frac{1}{2}\log\left|\bm{R}^{-1}(\bm{I}_n-\bm{S})\right|_+,
\end{eqnarray}
up to an additive constant that is independent of the parameters of interest. With this choice of $a$ and $b$ the posterior mean for $\sigma^2$ becomes
\begin{equation}\label{def:variance_estimator}
\hat\sigma^2 = \frac{\bm Y^T \bm R^{-1}({\bm I_n} - \bm S)\bm Y+1}{n+1},
\end{equation}
which is our estimator for the variance of the noise.

\subsection{Estimating Equations and Algorithm}\label{sec:EE}

The log-likelihood~\eqref{eq:RES} depends on the unknown parameters via the matrix $\bm{R}^{-1}(\bm{I}_n-\bm{S})$ only and
can be represented in a more convenient way as
$$
\bm{R}^{-1}(\bm{I}_n-\bm{S})=\bm\Phi_q\left\{\bm{I}_n+\mbox{diag}(\lambda n\bm{\eta}_q)\bm{\Phi}_q^T\bm{R}\bm{\Phi}_q\right\}^{-1}\mbox{diag}(\lambda n\bm{\eta}_q)\bm{\Phi}_q^T,
$$
using the Demmler-Reinsch basis; cf.~\ref{app:DRB}. The key result that allows us to handle efficiently the simultaneous estimation of $f$ and $r$ (or, equivalently, $\bm R$) is given in the following lemma whose proof can be found in~\ref{app:proofs:lemmas:diag2}, 

\begin{lemma}\label{lemma:diag2}
Let $\bm{R}$ be a $n\times n$ covariance matrix of a stationary process with spectral density $\rho$. Set $\rho_j=\rho(\pi t_j)$.
\begin{itemize}
\item[(i)]If $\rho$ belongs to the Hölder space $\mathcal{C}^{\gamma,\alpha}$, $\gamma\in\mathbb{N}$, $0<\alpha \leq 1$, i.e. $\rho^{(\gamma)}$ is $\alpha$-H\"older continuous, then, for $i,j=q+1,\ldots,n,$
$$\{{\bm \Phi}_q^T\bm{R}{\bm\Phi}_q\}_{i,j}={\rho}_j\delta_{i,j}+\mathbb{I}\{|i-j|\mbox{  is even}\} \cdot
\begin{cases}  O(\log(n)\cdot n^{-\gamma-\alpha+1}), &1<\gamma+\alpha\leq 2,\\O(n^{-1}), &2<\gamma+\alpha.\end{cases}$$
\item[(ii)] If $\rho$ belongs to the Sobolev space $\mathcal{W}_\beta(M)$, then, for $i,j=q+1,\ldots,n,$
$$\{{\bm \Phi}_q^T\bm{R}{\bm\Phi}_q\}_{i,j}={\rho}_j\delta_{i,j}+\mathbb{I}\{|i-j|\mbox{  is even}\} \cdot
\begin{cases} O(\log(n)\cdot n^{-1}), &\beta=2,\\O(n^{-1}), &\beta=3,4,\ldots.\end{cases}$$
\end{itemize}
\end{lemma}
\noindent (Note that the $O(n^{-1})$ term is uniform over $i,j$.)

That is, the Demmler-Reinsch basis $\bm{\Phi}_q$ asymptotically diagonalises $\bm{R}$ for any $q$ and the log-likelihood (\ref{eq:RES}) can thus be represented as
\begin{equation}\label{eq:app.like}
\begin{aligned}
\ell_n(\lambda,q,\bm{R})&=-\frac{n+1}{2}\log\left[ \sum_{i=q+1}^n\frac{B_{i}^2\lambda n\eta_{q,i}}{1+\lambda n\eta_{q,i} \rho_i }\{1+o(1)\}+O_P(1)\right]\\
&+\frac{1}{2}\sum_{i=q+1}^n\log\left(\frac{\lambda n\eta_{q,i}}{1+\lambda n\eta_{q,i}\rho_i}\right)\{1+o(1)\},
\end{aligned}
\end{equation}
where $B_{i}=\{\bm{\Phi}_q^T\bm{Y}\}_i$, see~\ref{app:likelihood}.
The representation in~\eqref{eq:app.like} should hold under wider generality but  Lemma~\ref{lemma:diag2} already provide a rather large model for the covariance structure of the noise. 

Since $\eta_{q,i}=[\pi\{i-(q+1)/2\}]^{2q}\{1+o(1)\}$ (see~\ref{app:DRB} for more details), it is straightforward to differentiate the approximative log-likelihood (\ref{eq:app.like}) w.r.t. $\lambda$, $q$ and $\rho_i$ to obtain after appropriate scaling the estimating equations
 \begin{align*}
T_\lambda(\lambda,q,\bm\rho) &= \left\{
\sum_{i=q+1}^n\frac{B_i^2 \lambda n\eta_{q,i}}{\big(1+\lambda n\eta_{q,i}\rho_i\big)^2} - \hat\sigma^2
\sum_{i=q+1}^n\frac1{1+\lambda n\eta_{q,i}\rho_i}
\right\}\{1+o(1)\},\\
T_q(\lambda,q,\bm\rho) &=\left\{
\sum_{i=q+1}^n\frac{B_i^2 \lambda n\eta_{q,i}\log(n\eta_{q,i})}{\big(1+\lambda n\eta_{q,i}\rho_i\big)^2} - \hat\sigma^2
\sum_{i=q+1}^n\frac{\log\big(n\eta_{q,i}\big)}{1+\lambda n\eta_{q,i}\rho_i}
\right\}\{1+o(1)\},\\
T_{\rho_i}(\lambda,q,\bm\rho) &=
\left(\frac{B_i^2 \lambda n\eta_{q,i}\rho_i}{1+\lambda n\eta_{q,i}\rho_i} - \rho_i\right)\{1+o(1)\},
\quad i=1, \dots, n,
\end{align*}
where
$$
\hat\sigma^2=\frac1{n+1}
\left(\sum_{i=q+1}^n\frac{B_i^2 \lambda n\eta_{q,i} }{1+\lambda n\eta_{q,i}\rho_i}+1\right).
$$

We would like to solve these equations simultaneously over $\lambda>0$, $\rho_i>0$, $i=1,\dots,n$, and $q\in\mathbb{N}$, to obtain joint estimates for all unknown model parameters.
However, this cannot be done directly due to interdependences and non-linearities in the equations.
In practice we proceed by solving the equations iteratively.
For each fixed $q$ we start with an initial guess $\hat{\bm{R}\,}^{(0)}$ (typically just an identity matrix), obtain a preliminary estimate $\hat{\lambda}^{(0)}$ and iterate to get $\hat\lambda_q$ and $\hat{\bm{R}\,}_q$. Finally, $q$ is chosen to solve $T_q(\hat\lambda_q,q,\hat{\,\bm{R}_q})=0$.

Since $\rho_i=\rho(\pi t_i)$ are values of a smooth function $\rho$ at given points, estimation of $\rho$ should be carried out over a space of smooth functions. In the Bayesian framework this can, conceivably, be accomplished by introducing a suitable prior on $\bm R$, which acts as a penalty term. 
For simplicity, we perform a post-processing procedure instead. First, $\tilde\rho_i$ are obtained as solutions of the corresponding estimating equations. Second, $\tilde{\rho}_i$ are smoothed, i.e., $\hat{\bm\rho\,}={\bm{S}}_{\xi,p,\bm{I}}\tilde{\bm{\rho}\,}$, where ${\bm{S}}_{\xi,p,{\bm{I}}}$ is a smoother matrix (\ref{eq:SR}) with parameters $\xi$, $p$ and $\bm{I}_n$.
(Note that this is inevitable: the estimates $\tilde{\rho}_i$ are inconsistent.)

All together, for each fixed $q$, at each step $\hat\lambda^{(j)}$ and $\tilde\rho_i^{(j)}$ are obtained as solutions of the corresponding estimating equations and $\tilde\rho_i^{(j)}$ are smoothed to get $\hat\rho_i^{(j)}$. The algorithm is iterated until convergence of $\hat\lambda$ and $\hat{\bm{\rho}}$, which are then used to get $\hat{q}$. Finally, $\hat{\bm{R}\,}$ is recovered from $\hat{\bm{\rho}\,}$ by the discrete Fourier transform; for the details see~\ref{app:aux:diagonalisation}. The summary of the estimation procedure is given in Algorithm~\ref{alg:estimation}:

\begin{algorithm}[!ht]
\caption{Estimation procedure.}
\For{$q$ in $Q_n$}{
set $j=1$ and $\hat{\bm{\rho}\,}^{(0)}\equiv1$\\
	\While{stopping criterium not met}{
		set $\hat \lambda^{(j)}$ to a solution of $T_\lambda(\lambda, q, \hat{\bm\rho}^{(j-1)})=0$ \;
          set $\tilde{\rho_i}^{(j)}$ to a solution of $T_{\rho_i}(\lambda^{(j)}, q, \hat{\rho_i}^{(j-1)})=0$\;
		compute $\hat{\bm \rho\,}^{(j)}$ by smoothing $\tilde{\bm\rho\,}^{(j)}$\;
    set $j=j+1$\;
	}
  set $\hat\lambda_q = \hat\lambda^{(j)}$ and $\hat{\bm\rho\,}_q = \hat{\bm\rho}^{(j)}$\;
}
set $\hat q$ to a solution of $T_q(\hat\lambda_q, q, \hat{\bm \rho\,}_q)=0$ over $Q_n$\;
set $\hat\lambda$ to $\hat\lambda_{\hat q}$\;
set $\hat{\bm \rho\,} = \hat{\bm \rho\,}_{\hat q}$\;
set $(\hat{\bm R\,})_{i,j} = \hat r_{|i-j|}$, with $\hat r_k=n^{-1}\sum_{l=1}^n\cos\left(k\pi\{l-1\}/\{n-1\}\right)\hat\rho_l$\;
\ \par
\label{alg:estimation}
\end{algorithm}

Here, $Q_n$ denotes the collection of values for $q$ under consideration.
The stopping rule is standard:
after each iteration we compare the change in the value of the estimate of $\lambda$ and the norm of the change in the estimate of $\bm\rho$;
if these fall bellow a threshold, then we stop iterating. The algorithm is 
fast and typically converges after just a few iterations.

In summary, the procedure outlined in Algorithm~\ref{alg:estimation} provides an approximation for the optimisers of the likelihood~\eqref{eq:app.like} with the extra property that the estimate of the spectral density is smooth (which is not accomplished by simply maximising the likelihood.)

\medskip

The theoretical derivation of the convergence rate for $\hat f$ and $\hat{\bm{R}\,}$ is not a trivial task due to the interdependence of the two estimators and requires a separate treatment. 
However, as far as $\hat f$ is concerned, in light of Lemma~\ref{lemma:diag2} it is clear from~\eqref{eq:SR} that $\bm R$ has a scaling effect on the smoother:
as $n$ grows $\bm S_{\bm I, \lambda/\delta} \le \bm S_{\bm R, \lambda} \le \bm S_{\bm I, \delta\lambda}$ for any $\bm R$ with eigenvalues on $[\delta,1/\delta]$, where $\bm A \le \bm B$ means that $\bm B-\bm A$ is positive semi-definite.
The conclusion is that the effect of the smoother $S_{\bm R, \lambda}$ is comparable to that of $S_{\bm I, \lambda}$ for a difference choice of $\lambda$.
We stress, however, that the small sample behaviour of the smoother $S_{\bm R, \lambda}$ is far superior in the presence of correlation; 
the smoother $S_{\bm I, \lambda}$ is well known to underperform particularly for positively correlated data since, contrary to the smoother $S_{\bm R, \lambda}$, it does not take the correlation structure of the noise into consideration.
From the likelihood in~\eqref{eq:app.like} it is clear, however, that the presence of correlation simply results in the presence of the (bounded) $\rho_i$ in the denominators.
This means that it should be possible to study the asymptotics of our estimator $\hat f$ in a similar way as when there is no correlation in the data as in~\cite{serra2017adaptive}.


Also the asymptotic behaviour of the estimator $\hat q$ should also be similar to when there is not correlation in the data; cf.~\cite{serra2017adaptive}.
Either way, the estimator should perform well even if $q$ is set by the user and not selected from the data; $q=2$ is a popular choice. 
The estimator for $\sigma^2$ should also be consistent.

In conclusion, the presence of short range dependent noise is close enough to the independent noise case to lead to the same large sample behaviour for estimators.
However, the small sample behaviour of estimators that ignore correlation can be significantly improved by use of our approach.

\subsection{Credible Sets}\label{sec:estimators:coverage}

Once estimates $\hat\lambda$, $\hat{\bm R\,}$, $\hat q$, and $\hat\sigma^2$ for respectively $\lambda$, $\bm{R}$, $\beta$, and $\sigma^2$ are available, these can be plugged into the marginal posterior for $\bm f$ to obtain the so called empirical, marginal posterior for $\bm f$:
\begin{equation}\label{eq:empirical_posterior}
\begin{aligned}
\hat\Pi^{\bm f}\!\big(\,\cdot \mid\bm Y\big) &=
\Pi^{\bm f}\!\big(\,\cdot \mid\bm Y,\hat\sigma^2,\hat\lambda,\hat q,\hat{\bm R\,}\big) \\ &=
\Pi^{\bm f}\!\big(\,\cdot \mid\bm Y,\sigma^2,\lambda,q,\bm R\big)\big|_{\big(\lambda,q,\bm R,\sigma^2\big)=\big(\hat\lambda,\hat q,\hat{\bm R\,},\hat\sigma^2\big)}.
\end{aligned}
\end{equation}
Given $\bm Y$, this is just a $t_{n+1}
\big(\hat{\bm f\,\,}, \hat\sigma^2\hat{\bm S\,}\hat{\bm R\,}\big)
$ distribution, which is centred at the spline estimate $\hat {\bm f\,}$, and whose covariance matrix depends on the (random) smoother $\hat{\bm S\,}$ -- which is just the smoother $\bm S$ with $(\hat\lambda,\hat q,\hat{\bm R\,})$ plugged in for $(\lambda,q,\bm R)$ -- and the estimates $\hat\sigma^2$ and $\hat{\bm R\,}$.
From this we can easily construct a credible set for the regression function.

In~\cite[Theorem 3]{serra2017adaptive}, it is shown that in the case where $\bm{R} = {\bm I_n}$ this set has two important properties if $L$ is taken appropriately large: 
a) the set contains the true underlying regression function with probability converging to 1, uniformly over a large subset of functions, and
b) with probability converging to 1, the (random) radius of $\hat C_n(L)$ is of the order of the minimax risk corresponding to the smoothness class to which the regression function belongs.

As argued in the previous section, the presence of correlation should have a relatively simple scaling effect on the smoothing parameter.
As such, the theoretical results of~\cite{serra2017adaptive} can in principle be extended to cover the correlated noise case as well, by (eventually) picking larger values of the multiplier $L$, but this problem will be studied in more generality elsewhere.
In this paper, we focus instead on the implementation, and numerical aspects of the procedure.
See for instance~\cite{sniekers2015adaptive,serra2017adaptive,rousseau2016asymptotic,yoo2016supremum} for more details on the use of credible sets as confidence sets, albeit for independent noise.

\section{Numerical Simulations}\label{sec:numerical}

In this section we investigate the small sample performance of our estimation procedure and compare it to several alternatives. Our simulation setup is as follows.
We consider two regression functions
\beqn
f_1(t)&=&\sum_{i=3}^n\psi_{3,i}(t)\{\pi(i-1)\}^{-3}\cos(2i)  \\
f_2(t)&=&2\sin(4\pi t), \phantom{\sum_{i=4}^n}
\eeqn
where $\psi_{3,i}$  is the $i$-th Demmler-Reinsch basis of ${\cal{W}}_3$ given explicitly in~\ref{app:DRB}, $n=800$ and $t\in[0,1]$. Both functions are subsequently scaled to have standard deviation $1$. The standard deviation of the residuals is taken to be $\sigma=0.33$ to imply a medium signal-to-noise ratio of $3$. All reported results are based on the Monte Carlo sample $M=500$. We consider $5$ types of the residual processes: an AR(1) process with the parameters $\phi=0.5$ and $\phi=0.9$, an ARMA$(2,2)$ process with $\phi=(0.7,-0.4)$ and $\theta=(-0.2,0.2)$ and two zero-mean Gaussian process with the correlation matrices $\bm{{R}}_1$ and $\bm{{R}}_2$ with $(i,j)$ entries given by $\cos(6.5 j)\exp(-|i-j|/10)$ and $\sin(1.5 j)\exp(-|i-j|/10)/(1.5j)$, respectively.
\begin{figure}[!ht]
\begin{tabular}{cc}
\includegraphics[width=0.45\textwidth]{"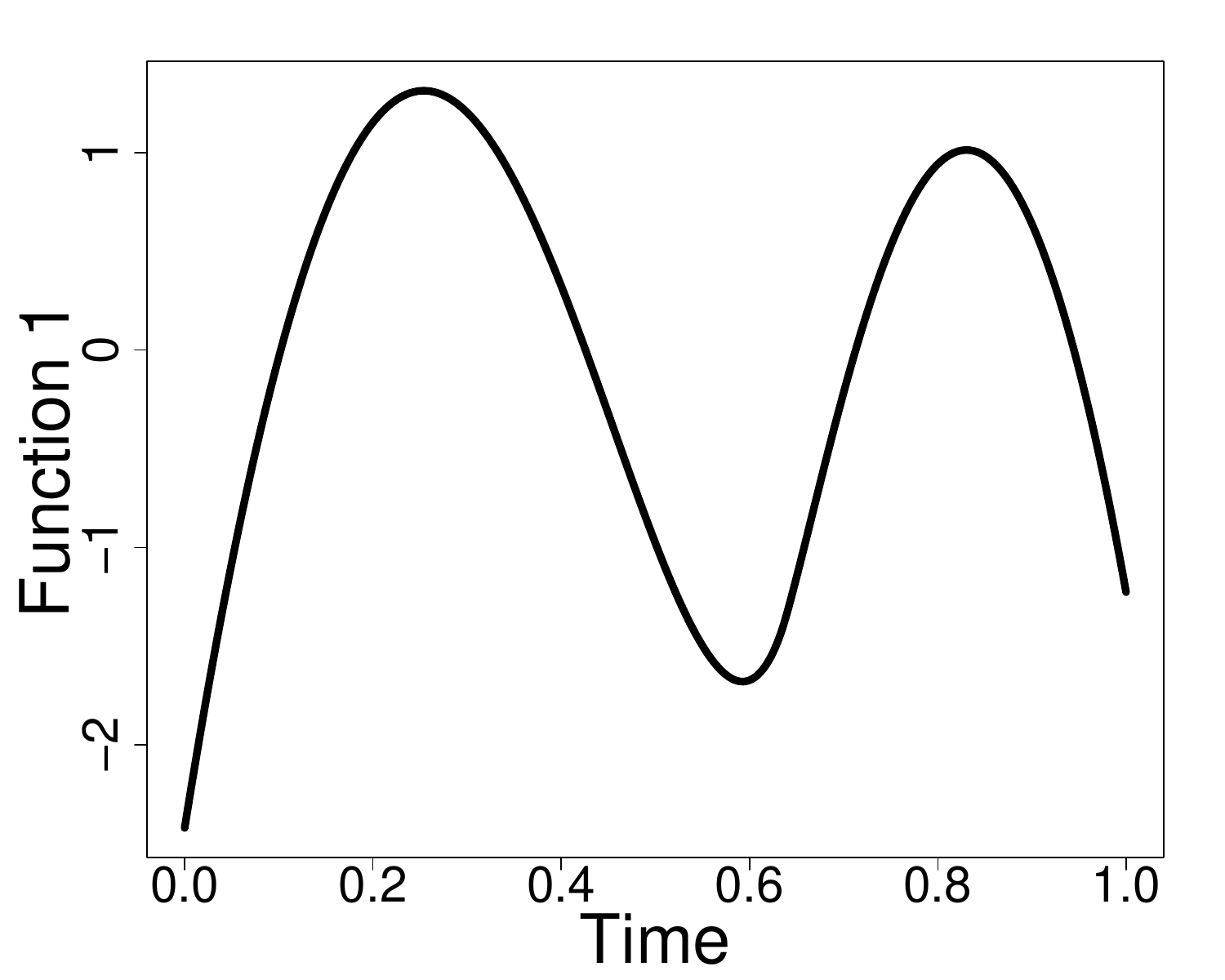"}&
\includegraphics[width=0.45\textwidth]{"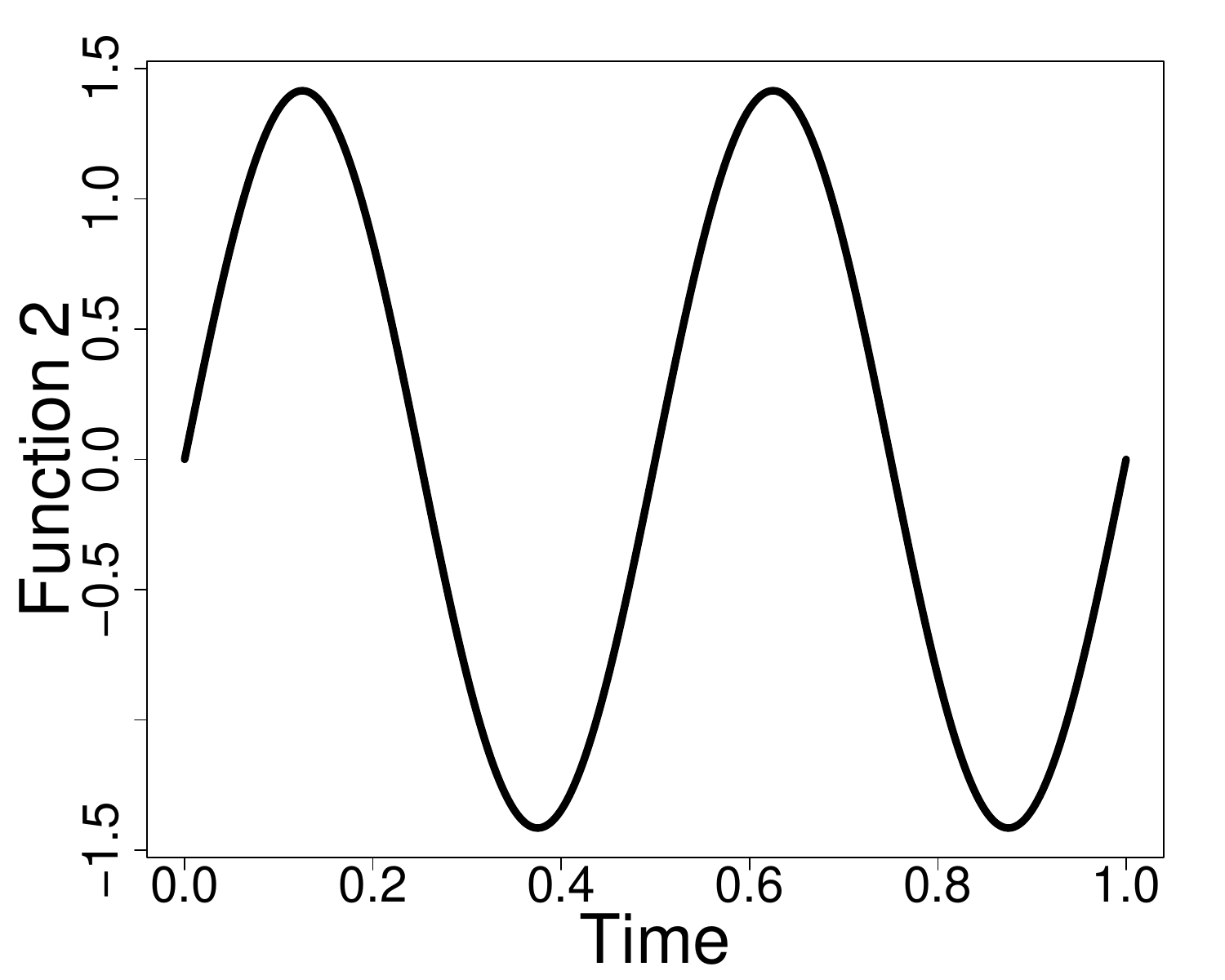"}\\
\includegraphics[width=0.45\textwidth]{"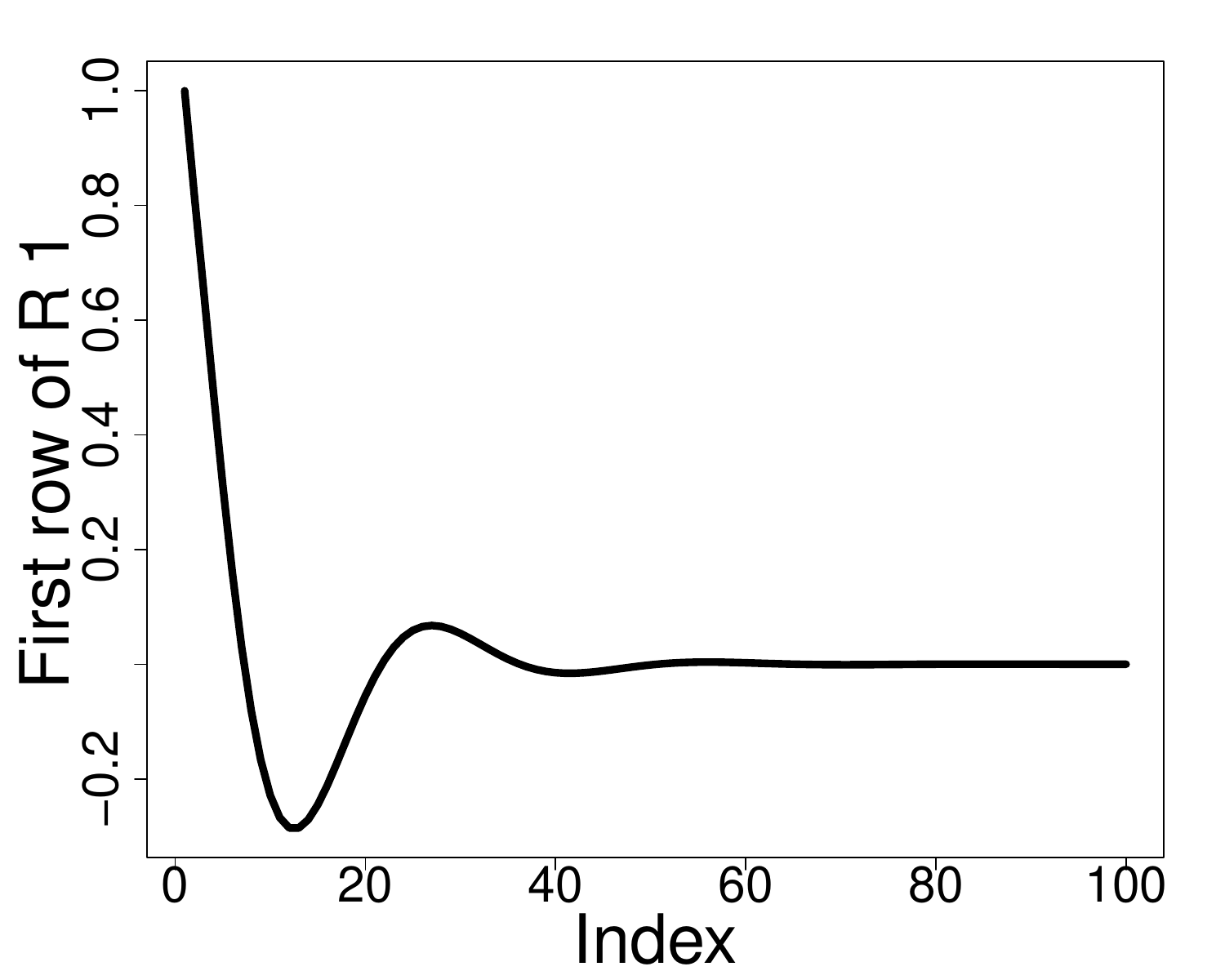"}&
\includegraphics[width=0.45\textwidth]{"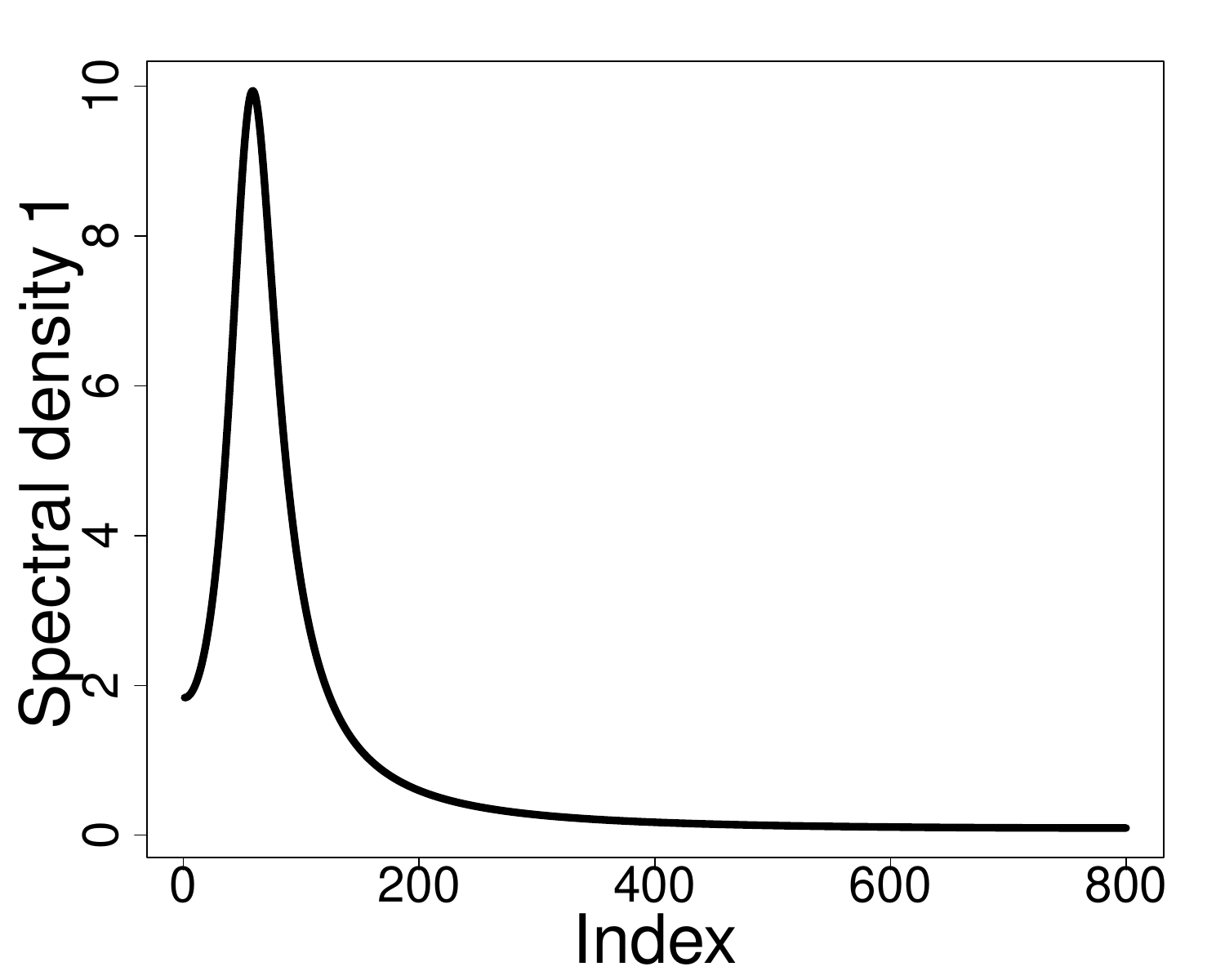"}\\
\includegraphics[width=0.45\textwidth]{"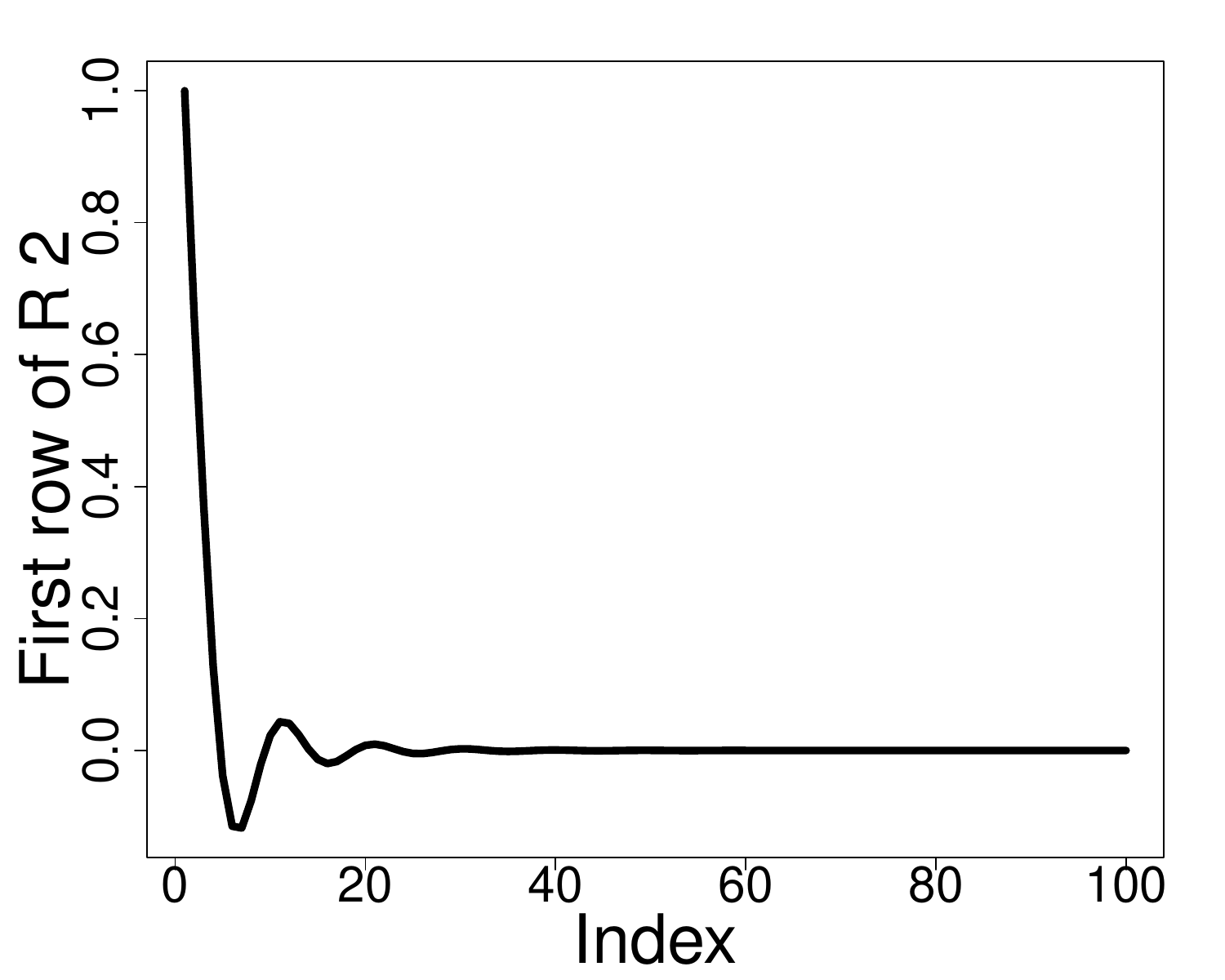"}&
\includegraphics[width=0.45\textwidth]{"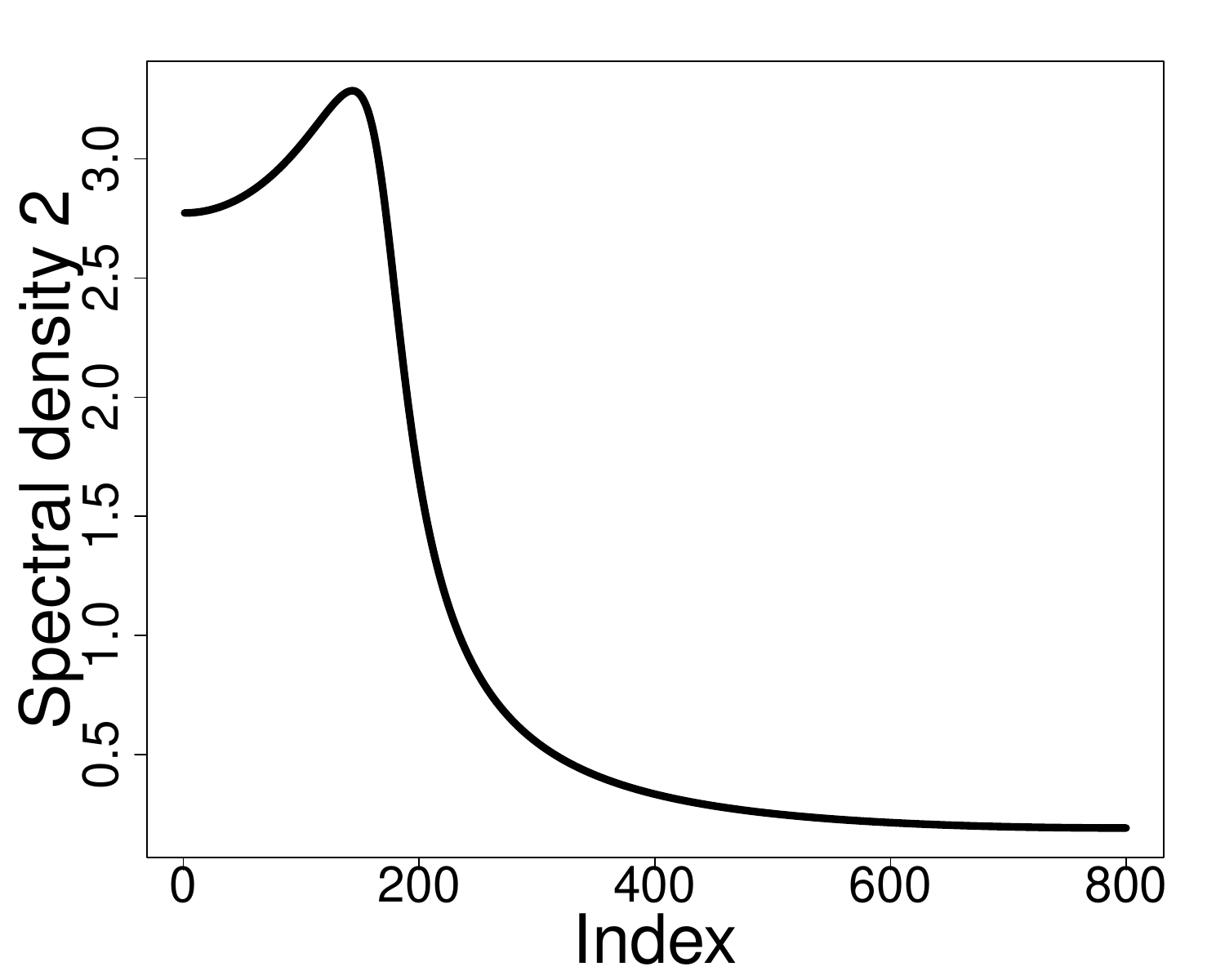"}\\
\end{tabular}
\caption{\small Regressions functions $f_1$ (top left), $f_2$ (top right), first $100$ values of the first row of ${\bm{{R}}}_1$ (middle left) with the corresponding spectral density (middle right) and of the first row of ${\bm{{R}}}_2$ (bottom left) with the corresponding spectral density.}
\label{fig:fun}
\end{figure}

The plots in the top row of Figure \ref{fig:fun} show both regression functions. The middle and bottom rows show the first $100$ elements of the first row of $\bm{{R}}_1$ (middle, left) and $\bm{{R}}_2$ (bottom, left) and the corresponding spectral densities (middle and bottom  right).

\begin{figure}[!ht]
\begin{tabular}{cc}
\includegraphics[width=0.45\textwidth]{"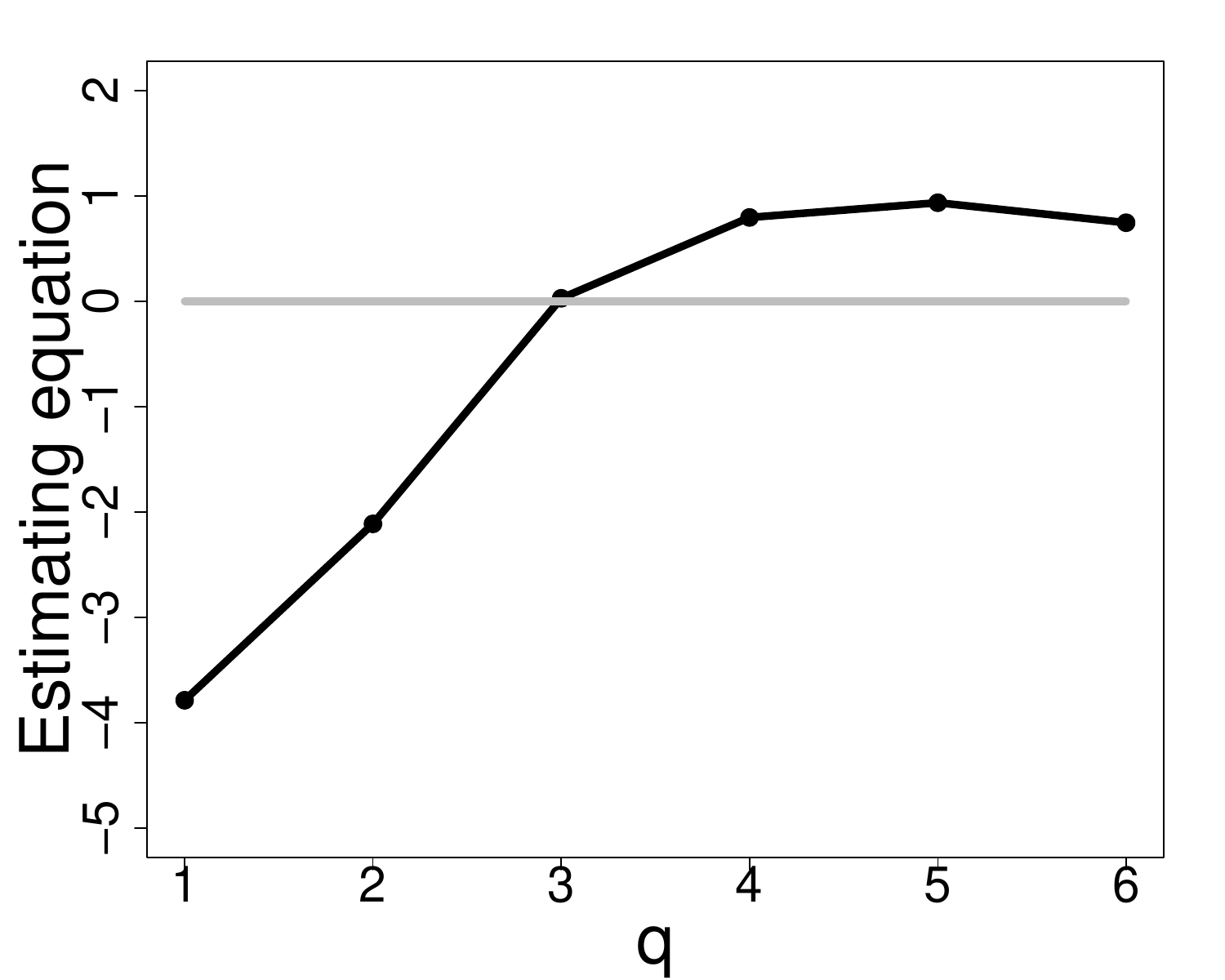"}&
\includegraphics[width=0.51\textwidth]{"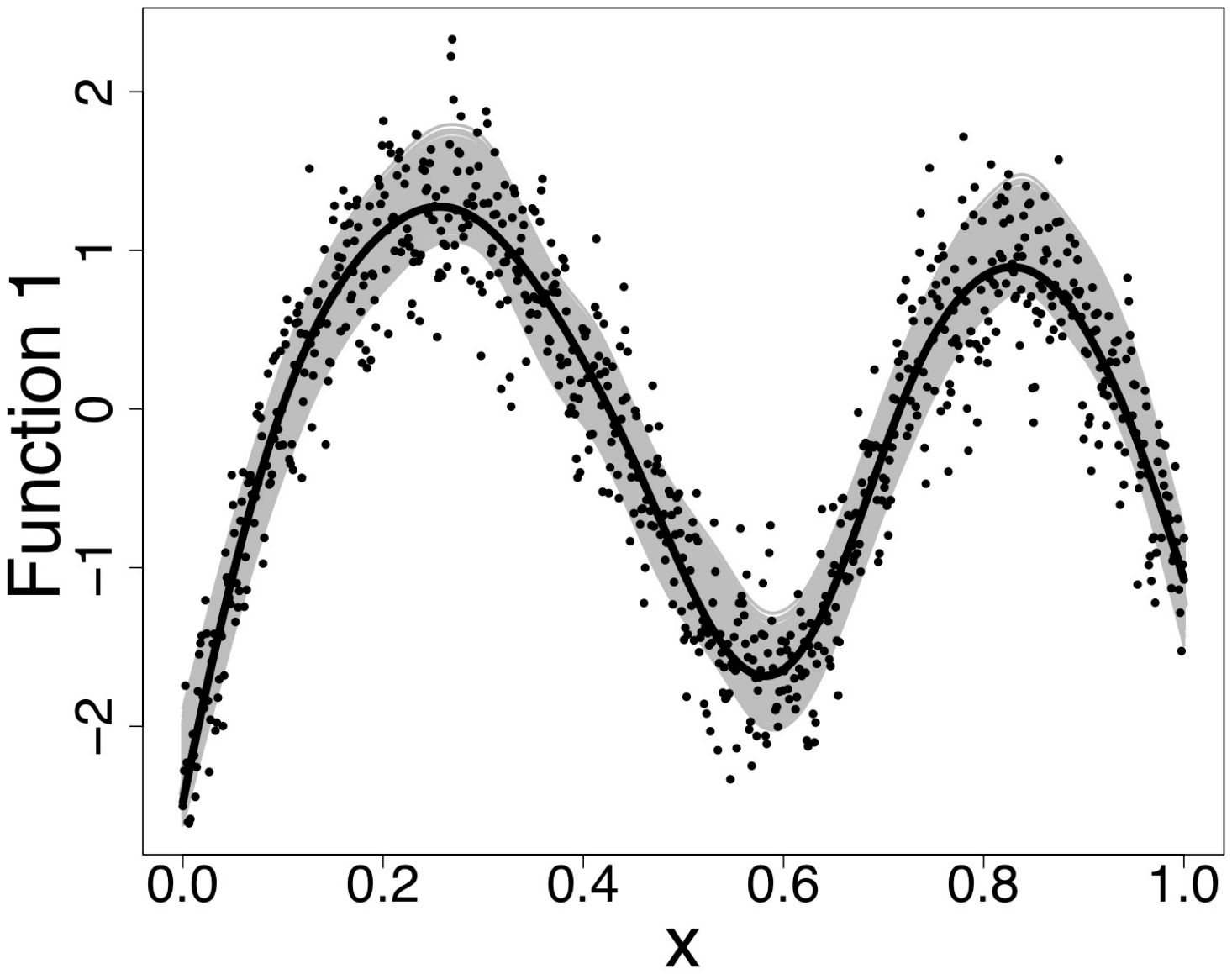"}\\
\includegraphics[width=0.45\textwidth]{"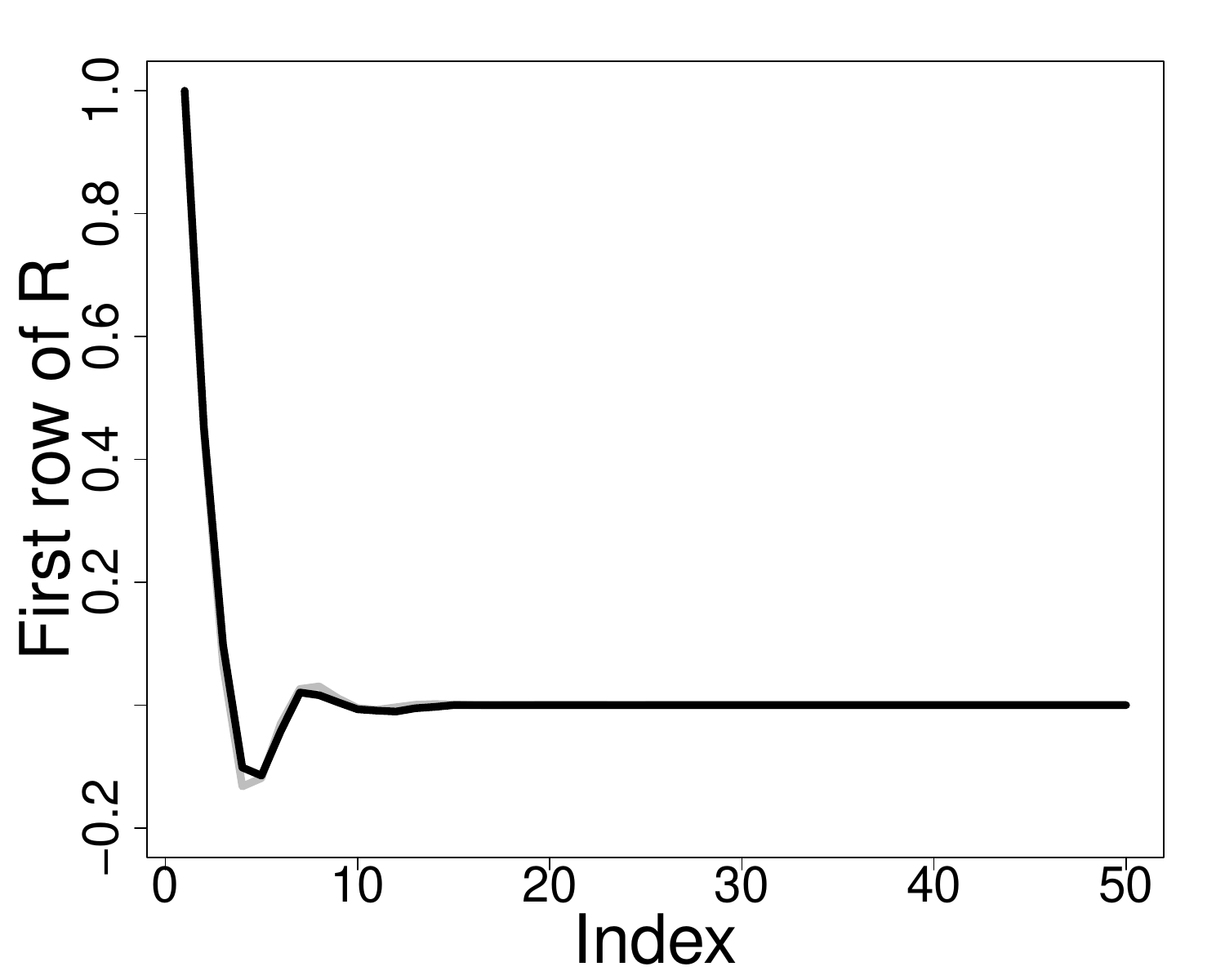"}&
\includegraphics[width=0.45\textwidth]{"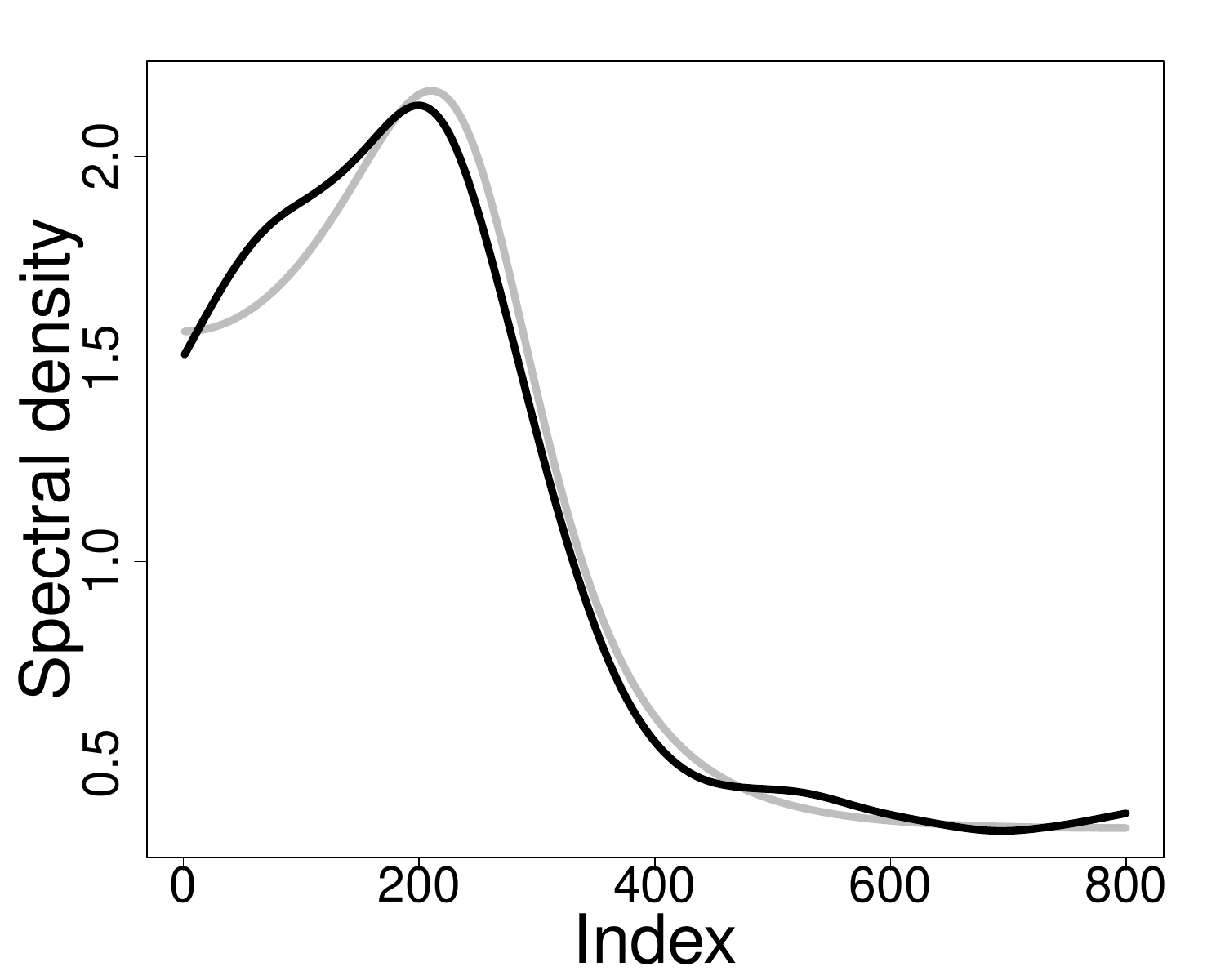"}\\
\end{tabular}
\caption{\small Estimating equation $T_q$ in the top left and  the corresponding estimator of $f_1$ (black line) together with the data (black points) and point-wise 95\% confidence intervals (grey area) in the top right. First $50$ elements of the first row of the correlation matrix in grey with its estimator in black (bottom left) and the true spectral density in grey with its estimator in black (bottom right).}
\label{fig:eBsc}
\end{figure}

Before we summarise the simulation results, we demonstrate how our method works in practice.
The data are simulated as described above with the regression function $f_1\in {\cal W}_3$ and ARMA$(2,2)$ dependence in the residuals.
After Algorithm~\ref{alg:estimation} has converged, one checks the estimate for $q$.
The top left plot of Figure \ref{fig:eBsc} shows the estimating equation $T_q(\hat\lambda,q,\hat{\bm{R}\,})$ which has a zero very close to $q=3$, as it should for $f_1\in{\cal{W}}_3$.
Next, the corresponding estimate for $f$ is obtained; it is shown in the top right plot of Figure~\ref{fig:eBsc}, together with the data (black dots) and 95\% confidence bands (grey area) constructed as described in Section~\ref{sec:estimators:coverage} (we used $L=\log n$.)
The estimate for ${{R}}_{1,j}$, $j=1,\ldots,50$ is shown in the bottom left plot of the same figure in black, which nearly coincides with the true autocorrelations, which are shown in grey.
This estimate was reconstructed by the discrete Fourier transform of the estimate of the spectral density, shown in black in the bottom right plot of Figure~\ref{fig:eBsc}.
The true spectral density is shown in the same plot in grey. Hence, our method allows for simultaneous, fully automatic and nonparametric estimation of $f$ (with data driven $q$, $\lambda$) and ${\bm{{R}}}$.

To the best of our knowledge, there is no other fully nonparametric method for joint estimation of mean and auto-covariance available. 
Nonetheless, we consider several approaches that can be at least partly compared to our method.
Among approaches that make a parametric assumption on $\bm{ R}$ we consider the well-established method based on splines that is implemented in the statistical software R in the function {\tt gamm} of package {\sl mgcv}, see \cite{Wood:2017}. We apply this estimation procedure  (further denoted by GAM) to the processes with parametric structure of $\{\epsilon(t_i)\}_{i=1}^n$ only and set the correlation structure parameters in {\tt gamm} to the truth, which is not available in practice.

Among approaches that make no parametric assumption on $\bm{ R}$ we consider the method of~\cite{Herrmann:1992}, subsequently referred to as HER. This kernel based method uses sample autocorrelation estimators to improve bandwidth selection and is developed under assumption of $m$-dependence in the residuals. However, the authors claim that the method works well for estimation of $f$ if the residuals satisfy ``some mixing conditions''. In general, this approach focuses on estimation of $f$ and the quality of the estimator for $r$ is not discussed.

Finally, we compare our estimator for the auto-covariance with the standard banded nonparametric estimator (further BAND). Since this approach requires a known mean, we plug in the true mean $f$ into the estimator, calculating
\beqn
\hat{R}_{i,j}(b)&=&\hat{r}_{|i-j|}\mathbb{I}(|i-j|\leq b),\;i,j=1,\ldots,n\\
\hat{r}_k&=&\frac{1}{n}\sum_{i=1}^{n-k}\{Y(t_i)-f(t_i)\}\{Y(t_{i+k})-f(t_{i+k})\},\; k=0,1,2,\ldots,
\eeqn
where $b$ is the banding parameter that needs to be chosen.

All these methods require selection of certain parameters. We proceed as follows. In the function {\tt gamm} we used low-rank splines with number of knots $n/4$,  B-spline basis of degree $3$, penalisation order $q=2$ and specified the correlation structure according to the true dependence structure in the residuals for the first $4$ residual processes.

For the method of \cite{Herrmann:1992} the parameter $m$ is set according to method (i) described in that paper (p.\ 787). Namely, we chose $m$ to be the largest integer such that $\hat{h}_m\geq 6/5\; \hat{h}_{m-1}$ and $m\leq 0.2\sqrt{n}$, where $\hat{h}_{m}$ is a selected bandwidth with the parameter $m$. This parameter $m$ is linked to the assumption of $m$-dependence in the residuals. In our experiments we noticed that the influence of $m$ on the estimator of $f$ is not very pronounced, but it does affect the estimator of the auto-covariance quite strongly. There is no simple data-driven approach to choose $m$ such that both mean and auto-covariance estimators are optimal in some sense. In our implementation we used function {\tt glkerns} of package {\sl lokern} by Eva Herrmann for the estimation of $f$ with the second order kernel and for the estimation of $f^{''}$ with the fourth order kernel.

Since there is no fully data-driven approach to banding parameter $b$ selection for the nonparametric covariance matrix estimation, we use an oracle band width, that is, such $b$ that minimises the empirical version of
$
\mathbb{E}\|\hat{\bm R}(b)-{\bm{{R}}}\|_\infty
$, where $\|A\|_\infty$ denotes the the maximum absolute row sum of matrix $A$; the calculation of the expectation is based on the Monte Carlo sample of size $500$.

Even though our approach is adaptive and $\beta$ can be estimated from the data, we set $q=2$ for our method to be able to compare mean estimators across all procedures, which should have the same convergence rate. All other parameters are estimated from the data. We refer to our method as BAS.

The results are summarised in Table \ref{table:results}. For all dependence structures and all functions we calculate
 \beqn
 A(\hat{f}_j)&=&\frac{1}{Mn}\sum_{k=1}^n\sum_{i=1}^M\{f_j(x_k)-\hat{f}_{j,i}(x_k)\}^2\\
 A(\hat{R}_j)&=&\frac{1}{Mn}\sum_{k=1}^n\sum_{i=1}^M\{{{R}}_{j}(k)-\hat{R}_{j,i}(k)\}^2,\;\;j=1,2.
 \eeqn
 Here $\hat{f}_{j,i}$ denotes an estimator of $f_j$ in $i$th Monte Carlo run with the residuals following one of the six processes. Similarly, ${{R}}_j(k)$ denotes the $k$-th entry of the first row of one of six true residual correlation matrices added to the $j$th regression function $f_j$ and $\hat{R}_{j,i}(k)$ is its estimator in the $i$th Monte Carlo run.

 \begin{table}[!ht]
   \centering
 \begin{tabular}{l|rrr|rrrr}
 \hline
    \multicolumn{7}{l}{$\bf{f_1}$}\\\hline
 &\multicolumn{3}{l|} {$A(\hat{f_1})$}&\multicolumn{4}{l} { $A(\hat{R}_1)$}\\\hline\hline
 Correlation&BAS&GAM&HER&BAS&GAM&HER&BAND\\\hline
  AR1(0.5)  & 6.992 & 6.866 & 6.433 & 0.019 & 0.004 & 0.017 & 0.012 \\
AR1(0.9)  &107.605 & 96.652 & 300.940 & 0.830 & 0.140 & 2.934 & 0.293 \\
ARMA(2,2)& 4.163 & 4.012 & 3.797 & 0.015 & 0.011 & 0.031 & 0.013  \\
GP1&  4.161 & -- & 3.727 & 0.200 & --& 0.789 & 0.124 \\
GP2 & 5.495 & -- & 4.771 & 0.047 & -- & 0.087 & 0.048  \\\hline\hline
    \multicolumn{7}{l}{$\bf{f_2}$}\\\hline
 &\multicolumn{3}{l|} {$A(\hat{f_2})$}&\multicolumn{4}{l} { $A(\hat{R}_2)$}\\\hline\hline
 Correlation&BAS&GAM&HER&BAS&GAM&HER&BAND\\\hline
AR1(0.5)  & 7.207 & 7.028 & 6.114 & 0.022 & 0.005 & 0.016 & 0.012   \\
AR1(0.9)  &121.442 & 113.355 & 308.155 & 0.848 & 0.163 & 2.885 & 0.271 \\
ARMA(2,2) & 4.417 & 4.236 & 3.482 & 0.016 & 0.012 & 0.033 & 0.014  \\
GP1 &4.449 &-- & 3.603 & 0.197 & --& 0.786 & 0.117\\
GP2 &5.769 & -- & 4.541 & 0.050 &-- & 0.090 & 0.046  \\\hline\hline
 \end{tabular}
 \caption{\small Simulation results. Values of $A(\hat{f_j})$ and $A(\hat{R}_j)$, $j=1,2$ are multiplied by $10^3$.}
 \label{table:results}
 \end{table}
Method HER performs best in estimation of the mean, except for the case with AR(1) error process with the parameter $0.9$, where this approach fails completely.  For estimation of  auto-covariances HER performs, in contrast, worst, which is most likely due to the ad-hoc choice of the tuning parameter $m$. For parametric structures of the error process method GAM performs very similar to BAS in estimation of the mean and is better in estimation of the auto-covariances. This is not surprising, since our nonparametric method should have a slower convergence rate than the parametric one. However, in simulations we used the correct parametric  model specification in the function {\tt gamm}, which is not available in practice. Finally, BAND's performance in nonparametric estimation of auto-covariances is very similar to BAS, even though BAND uses the known mean and oracle banding parameter, both of which are not available in practice. All together, our fully nonparametric and data-driven method is competitive to the considered alternatives, most of which rely on oracle choices for their parameters.

\section{Real Dataset}
\label{sec:examples}

We consider the data from the load forecasting track of the Global Energy Forecasting Competition 2012 (\url{http://www.drhongtao.com/gefcom/2012}). These are data on hourly loads of a US facility at $20$ zones from the 1st hour of January 1st, 2004 to the 6th hour of June 30th, 2008. The goal of the competition was to make a one week out-of-sample forecast for each of the 20 time series, as well as backcast certain missing values within the observational period. Here we are not interested in forecasting the data, but rather understanding their structure. We consider the mean over all $20$ zones over the whole time period -- all together $n=1650$ observations. Missing values were imputed using R package {\tt Hmisc}; omitting these missing values lead to the same estimators and same conclusions.

\begin{figure}[!ht]
\begin{center}
\includegraphics[width=0.9\textwidth]{"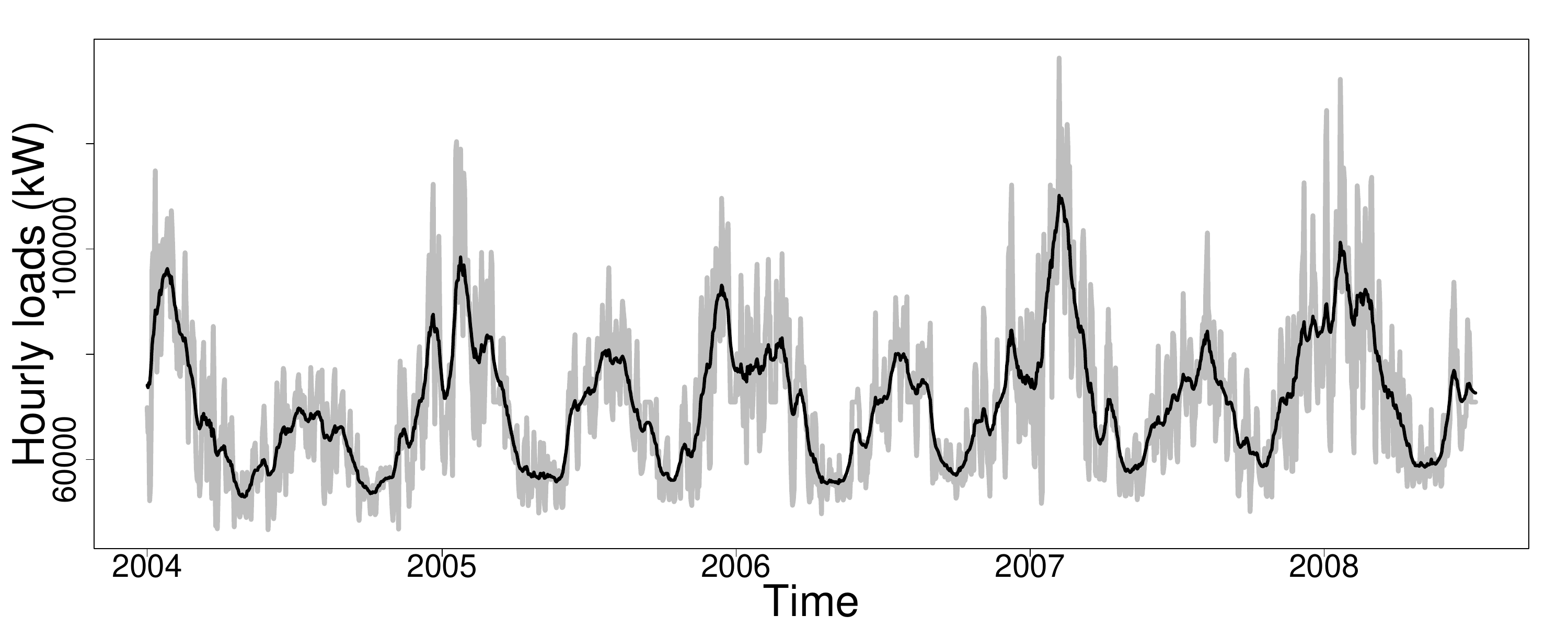"}
\includegraphics[width=0.45\textwidth]{"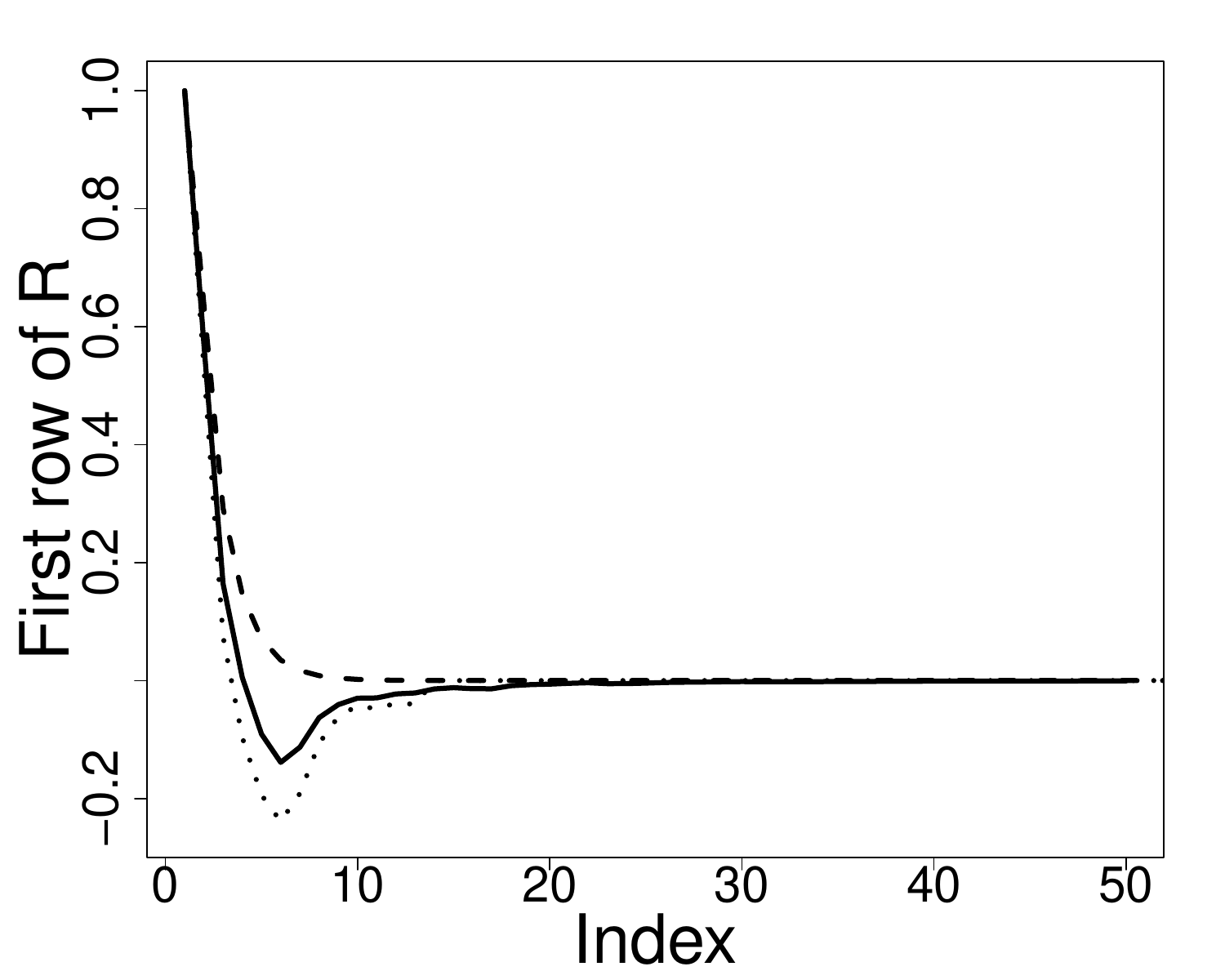"}
\includegraphics[width=0.45\textwidth]{"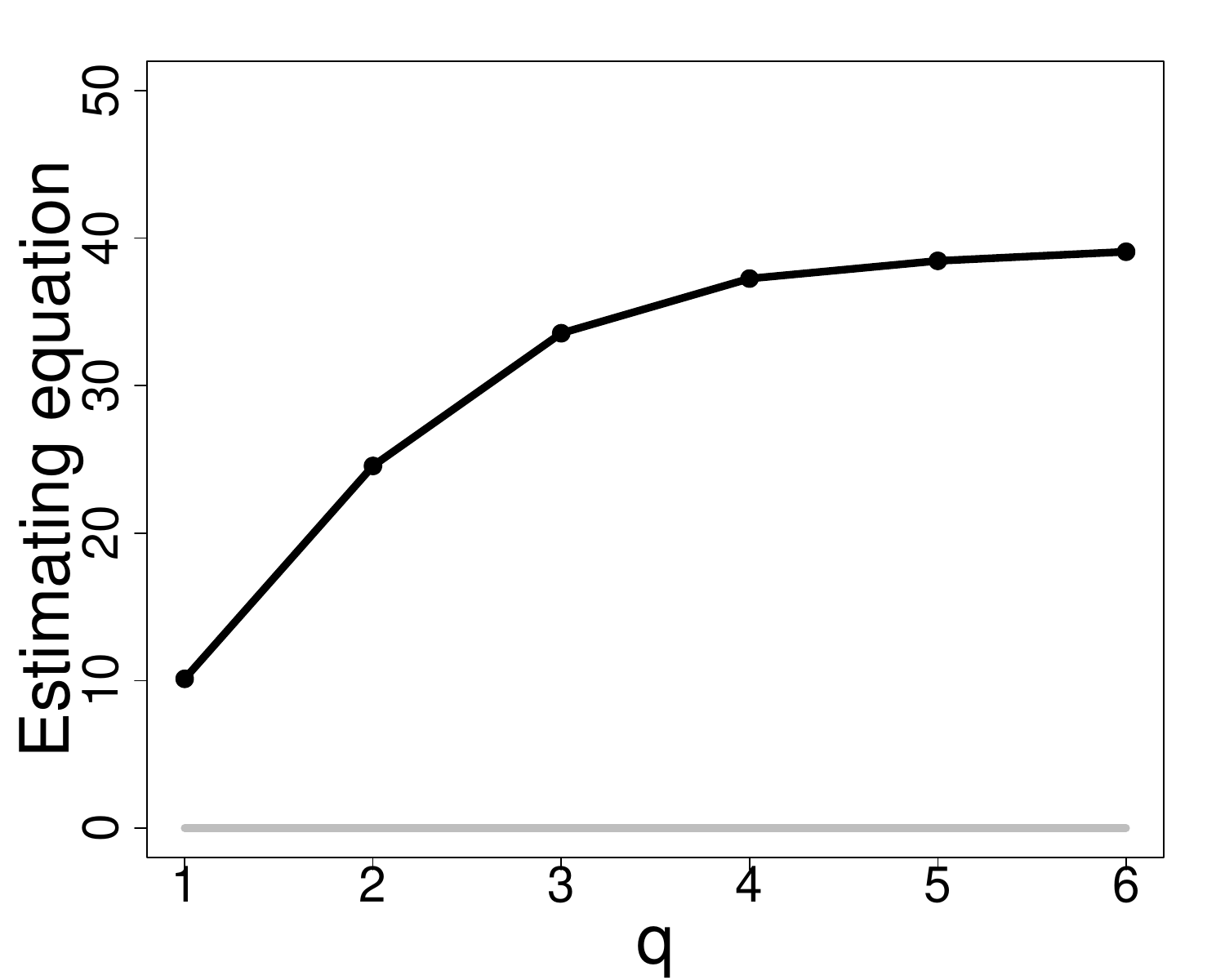"}
\end{center}
\caption{\small Estimators of the mean (top), first row of the correlation matrix (first 50 values, bottom left) and the estimating equation for the smoothness degree $q$ (bottom right) of the hourly loads of a US facility. Data are shown in grey, BAS estimators are the black solid lines, GAM estimators are dashed lines, BAND estimator for the covariance is the dotted line.}
\label{fig2}
\end{figure}

Fitting the data with our approach delivered a positive and increasing function in the estimating equation for $q$, which implies that the mean function has either $1$ continuous derivative or is even less smooth. This suggests that a fit with $q=1$ is more appropriate than a fit with $q=2$. This agrees with the nature of the data which is prone to display peak loads;
in the literature similar data are treated e.g., by a mixture of smoothing splines and wavelets, where the later pick up the peaks, see \cite{Amato:2017}. The estimator of the mean with $q=1$ is shown in the top plot of Figure~\ref{fig2} as a black line. The corresponding autocorrelations are shown as solid black line in the bottom left plot.

 Setting $q=1$ in GAM with an AR(2) process for the residuals results in the estimator of the mean that is very close to the one obtained with BAS, but having somewhat less pronounced peaks (not clearly visible in the plot). The corresponding autocorrelation estimator is shown as a dashed line in the bottom left plot of Figure \ref{fig2}. Assuming an ARMA$(2,2)$ process leads to nearly the same estimator. Since the HER method is defined only for even order kernels, we can not obtain the fit that would be comparable with the fit by BAS with $q = 1$.  Subtracting the estimated mean from the data we employed a banded nonparametric estimator BAND for the covariance with band $b=12$, which is shown as a dotted line in the bottom left plot and is closer to the BAS estimator of the covariance.

Estimation with BAS setting $q=2$ gives estimates of the mean that are very close to the one obtained by GAM, setting the correlation structure to an AR(1) process; this is shown in Figure~\ref{fig:intro}, right plot. The corresponding autocorrelation estimators are also reasonably close. The residual analysis suggests, however, that the fit with $q=1$ is more appropriate for this data.

\section{
Conclusions
}\label{sec:conclusions}

Correlation is ubiquitous in applications, particularly when data is collected sequentially over time, and ignoring this can have severe consequences for inference.
Covariance in data is typically taken into account by either making parametric assumptions on the dependence structure, or by relying on introducing tuning parameters that then have be set heuristically.
We propose a fully automatic, nonparametric method to estimate both mean function and auto-covariance function, and further supply credible sets for the mean function that quantify the uncertainty of the estimator. The order of the splines to be used in the estimator is also estimated from the data, fully data driven. The approach is implemented in the R package {\sl eBsc} (available from the authors), and delivers results in a quick and numerically stable way.

Derivation of the convergence rate turned out to be a non-trivial task due to the interdependence of the estimators and will be performed in a separate work. 
Nonetheless, for short range dependent noise models as covered here, we expect the results, both in terms of rates and uncertainty quantification to be similar to the independent noise case.
Since there are no competing fully non-parametric, fully data-driven, methods available in the literature, in our numerical simulations we compare our approach with methods from the literature that require knowledge of oracle parameters.
Even so, our numerical simulations suggest that our method is competitive to these alternatives.

\appendix
\section{Auxiliary Results}\label{app:aux}

In this appendix we collect a number of technical results that are used throughout this paper.

\subsection{Demmler-Reinsch Basis}\label{app:DRB}

Let $\{\psi_i\}_{i=1}^\infty$ denote the Demmler-Reinsch basis of
${\cal{W}}_\beta(M)$, such that
\begin{equation*}
\nu_{\beta,i}\int_0^1\psi_{\beta,i}(x)\psi_{\beta,j}(x)dx=\nu_{\beta,i}\delta_{ij}=\int_0^1\psi_{\beta,i}^{(\beta)}(x)\psi_{\beta,j}^{(\beta)}(x)dx.
\end{equation*}
For a precise definition of the space ${\cal{W}}_\beta(M)$, $\beta\in\mathbb{N}$, see~\cite{serra2017adaptive}.
\cite{Rosales2016} found explicit expressions for $\psi_{\beta,i}$ and $\nu_{\beta,i}$ as a solution to
\beqn
&&(-1)^q\psi_{\beta,i}^{(2\beta)}=\nu_{\beta,i}\psi_{\beta,i}\\
&&\psi_{\beta,i}^{(l)}(0)=\psi_{\beta,i}^{(l)}(1)=0,\;\;l=\beta,\beta+1,\ldots,2\beta-1.
\eeqn
In particular, $\nu_{\beta,1}=\cdots=\nu_{\beta,\beta}=0$ and
\beq
\nu_{\beta,i}&=&\left\{\pi\left(i-\frac{\beta+1}{2} \right)  \right\}^{2\beta},\;\;i=\beta+1,\beta+2,\ldots\nonumber\\
\psi_{\beta,i}(x)&=&\sqrt{2}\left[\cos\left\{\pi \left(i-\frac{\beta+1}{2}\right)x+\pi\,\frac{\beta-1}{4}\right\}+T_i(x)\right]\label{eq:DRSobolev}
\eeq
where
$$
T_i(x)=\sum_{a_j\in S(\beta)}r_j\left[ \exp\left\{ -a_j\pi \left(i-\frac{\beta+1}{2}\right)x\right\} +(-1)^{i+1}\exp\left\{ -a_j\pi \left(i-\frac{\beta+1}{2}\right)(1-x)\right\} \right],
$$
for $S(\beta)=\cup_j\left\{(-1)^{j/(2\beta)},\overline{(-1)^{j/(2\beta)}}\right\}$, with $0\leq j\leq \beta-2$ taking  odd values for $\beta$ odd and even values for $\beta$ even. Constants $r_j$ are known and depend on $\beta$ only. Note that $T_i(x)$ vanish exponentially fast away from the boundaries. \\
The Demmler-Reinsch basis of the natural spline space of degree $2q-1$
with knots at
observations ${\cal{S}}_{2q-1}(\bm{x})$ is uniquely defined via
\begin{equation*}
\eta_{q,i}\sum_{k=1}^n\phi_{q,i}(x_k)\phi_{q,j}(x_k)=\eta_{q,i}\delta_{ij}=\int_0^1\phi_{q,i}^{(q)}(x)\phi_{q,j}^{(q)}(x)dx
\end{equation*}
and $\bm{\Phi}_q=\bm{\Phi}_q(\bm{x}) = [\phi_{q,1}(\bm{x}), \dots,
\phi_{q,n}(\bm{x})] =
[\phi_{q,j}(x_i)]_{i,j=1}^n$ is the corresponding basis matrix. \\
\cite{Utreras1980} used the results of \cite{Fix1972} to show that $|n \eta_{q,i}-\nu_{q,i}|=O(n^{-2})$. From \cite{Fix1972} and \cite{Fix1973} also follows that $\|\sqrt{n}\phi_{q,i}-\psi_{q,i}\|_{{\cal{W}}_q}=O(n^{-1})$, or, equivalently, that  $\|\phi_{q,i}-\psi_{q,i}/\sqrt{n}\|_{L_2}=O(n^{-3/2})$.\\
To apply results by \cite{Fix1972} and \cite{Fix1973} the standard result \citep[see Lemma 3.2 in][]{Utreras1980} is used.
\begin{lemma}
\label{lemma:quadrature}
Let $f\in W_\beta(M)$, $\beta\geq 2$, then
$$
\left| \frac{1}{n}\sum_{i=1}^nf(t_i)-\int_0^1f(t)dt\right|\leq \frac{c}{n^2}\|f\|_{{\cal{W}}_\beta},
$$
where
$
t_i=(i-1)/(n-1)$, $i=1,\ldots,n$.
\end{lemma}

\subsection{Matrix Identities}\label{app:aux:matrix_idents}

Here, we derive some matrix identities.
The smoother matrix $\bm S$ satisfies
\[
\bm S = \bm{R}^{1/2}\left\{\bm{I}_n + \lambda\bm{R}^{1/2} \bm{\Phi}_q\diag(n\bm{\eta}_q)\bm{\Phi}_q^T\bm{R}^{1/2} \right\}^{-1}\bm{R}^{-1/2}.
\]
From this expression is it clear that $\bm R^{-1}\bm S = \bm S^T \bm R^{-1}$ so that in particular
\[
\bm R^{-1}({\bm I_n}-\bm S) = ({\bm I_n}-\bm S)^T \bm R^{-1}.
\]
From the definition of $\bm S$ is is also simple to check the scaling relation
\[
\bm R^{-1}(\bm S^{-1}-{\bm I_n}) = \lambda {\bm\Phi_q} \diag(n\bm\eta_q){\bm\Phi_q}^T = \bm S_{\bm I}^{-1}-\bm I_n.
\]

The estimating equation for $\lambda$ is obtained in a straightforward way by noting that since $\partial\bm S/\partial\lambda = -({\bm I_n} - \bm S)\bm S/\lambda$, then
\[
\frac{\partial \bm R^{-1}({\bm I_n}-\bm S)}{\partial \lambda} =
\frac1\lambda \bm R^{-1}({\bm I_n}-\bm S)\bm S,
\]
and the estimating equation for $q$ is based on the fact that
\begin{align*}
\frac{\partial \bm R^{-1}({\bm I_n}-\bm S)}{\partial q} &=
\bm R^{-1}\bm S\frac{\partial \bm S^{-1}}{\partial q}\bm S =
\frac\lambda q \bm R^{-1}\bm S\bm R \bm\Phi_q\diag\big\{n\bm\eta_{q}\circ\log(n\bm\eta_{q})\big\}\bm\Phi_q^T\bm S\\ &=
\frac\lambda q\bm S^T \bm\Phi_q\diag\big\{n\bm\eta_{q}\circ\log(n\bm\eta_{q})\big\}\bm\Phi_q^T\bm S\\ &=
\frac1 q\bm R^{-1}(\bm I-\bm S) \bm\Phi_q\diag\big\{\log(n\bm\eta_{q})\big\}\bm\Phi_q^T\bm S,
\end{align*}
where $\circ$ denotes the Hadamard product.
For the derivation of the estimating equations for the $\rho_i$ we use the fact that
\[
\frac{\partial \bm S}{\partial\rho_i} =
-\bm S\frac{\partial \bm S^{-1}}{\partial\rho_i}\bm S =
-\bm S\frac{\partial \bm R}{\partial\rho_i}(\bm S_{{\bm I_n}}^{-1}-{\bm I_n})\bm S =
-\bm S\frac{\partial \bm R}{\partial\rho_i}\bm R^{-1}({\bm I_n}-\bm S),
\]
such that, combining the above, we see that by using the chain rule
\begin{align*} &
\frac{\partial \bm R^{-1}(\bm I_n-\bm S)}{\partial \rho_i} =
-\bm R^{-1}\frac{\partial\bm R}{\partial\rho_i}\bm R^{-1}(\bm I_n -\bm S) + \bm R^{-1}\frac{\partial\bm S}{\partial\rho_i}\\ &\quad=
-\bm R^{-1}({\bm I_n} - \bm S)\frac{\partial\bm R}{\partial\rho_i}\bm R^{-1}({\bm I_n} -\bm S) =
-({\bm I_n} - \bm S)^T\bm R^{-1}\frac{\partial\bm R}{\partial\rho_i}\bm R^{-1}({\bm I_n} -\bm S).
\end{align*}

\subsection{$\bm R$ and its Spectral Density}\label{app:aux:diagonalisation}

Let $\bm R$ be a Toeplitz matrix in that $(\bm R)_{i,j} = R_{i,j} = r_{i-j}$ for some sequence $\{r_i\}_{i\in\mathbb{Z}}$.
Let us further assume that $R_{i,j}=0$, if $|i-j|>m$, $m\in\mathbb{N}$.
Denote the Fourier spectrum of $\bm R$ as
\[
\rho(x) = \sum_{k=-m}^m r_k \exp(\im k x),
\quad\text{so that}\quad
r_k = \frac1{2\pi} \int_{-\pi}^{\pi} \rho(x)\exp(-\im k x)\, dx,
\]
where $\im$ denotes the imaginary unit.
If $\bm R$ is the correlation matrix of a stationary process, then additionally we have that $r_{i-j}=r_{j-i}$ so that we can write
\[
\rho(x) = 1 + 2 \sum_{k=1}^m r_k \cos(k x)
\quad\text{and}\quad
r_k = \frac1{2\pi}\int_{-\pi}^\pi \cos(kx)\rho(x)\,dx = \int_0^1 \cos(k\pi x)\rho(\pi x)\,dx.
\]
In practice we approximate $r_k$ via
\[
r_k = \frac1n\sum_{i=1}^n\cos\left(k\pi\frac{i-1}{n-1}\right)\rho_i+O(n^{-1}),\quad\mbox{where}\;\;\rho_i = \rho\left(\pi\frac{i-1}{n-1}\right).
\]

\subsection{Proof of Lemma~\ref{lemma:diag2}}\label{app:proofs:lemmas:diag2}

$(i)$ W.l.o.g. assume that $n$ is even. Noting that $t_i=(i-1)/(n-1)$ and denoting   $t_{j,q}=\{j-(q+1)/2\}/(n-1)$ we get for $i,j=q+1,\ldots,n$
\beq
\begin{aligned}
\label{eq:psi}
\{{\bm \Phi}_q^T\bm{R}{\bm \Phi}_q\}_{i,j}&=\sum_{l=1}^n\phi_{q,i}(t_l)\sum_{k=1}^n\sqrt{\frac{2}{n}}\cos\{\pi(k{-}1)t_{j,q}{+}\pi(q{-}1)/4\}r_{|l-k|}\\
&{+}\sum_{l=1}^n\phi_{q,i}(t_l)\sum_{k=1}^n\left[\phi_{q,j}(t_k)-\sqrt{\frac{2}{n}}\cos\{\pi(k{-}1)t_{j,q}{+}\pi(q{-}1)/4\}\right]r_{|l-k|}\\
&=\rho(\pi t_{j,q})\delta_{i,j}\\
&{-}\{1{+}(-1)^{|i-j|}\}\sqrt{\frac{2}{n}}\sum_{l=1}^{n/2}\phi_{q,i}(t_l)\sum_{s=n-l+1}^{\infty}\cos\{\pi (l{-}1{+}s)t_{j,q}{+}\pi(q{-}1)/4\}r_{s}\\
&{-}\{1{+}(-1)^{|i-j|}\}\sqrt{\frac{2}{n}}\sum_{l=1}^{n/2}\phi_{q,i}(t_l)\sum_{s=l}^{\infty}\cos\{\pi(l{-}1{-}s)t_{j,q}{+}\pi(q{-}1)/4\}r_{s}\\
&{-}\rho(\pi t_{j,q})\sum_{l=1}^n\phi_{q,i}(t_l)\left[\phi_{q,j}(t_l)-\sqrt{\frac{2}{n}}\cos\{\pi(l{-}1)t_{j,q}{+}\pi(q{-}1)/4\}\right]\\
&{+}\sum_{l=1}^n\phi_{q,i}(t_l)\sum_{k=1}^n\left[\phi_{q,j}(t_k)-\sqrt{\frac{2}{n}}\cos\{\pi(k{-}1)t_{j,q}{+}\pi(q{-}1)/4\}\right]r_{|l-k|}.
\end{aligned}
\eeq

We used $\sum_{|s|\leq n/2}\cos(\pi s t_{j,q})r_s= \rho(\pi t_{j,q}) + \sum_{|s|>n/2}^\infty \cos(\pi s t_{j,q})r_s$, the cosine addition identity, and that cosine and sine are even and odd functions, respectively. Four last terms in (\ref{eq:psi}) vanish for $|i-j|$ odd; for the last two terms this follows since $\phi_{q,j}(t_k)$ has the same number of sign changes as $\cos\{\pi(k-1)t_{j,q}+\pi(q-1)/4\}$ (see~\ref{app:DRB}) and $\phi_{q,i}(t_k)$ is an even function for $i$ odd and odd function for $i$ even. If $|i-j|$ is even, then the second term in (\ref{eq:psi}) is of order $O(n^{-\gamma-\alpha+1})$ since $\phi_{q,i}(x)=O(n^{-1/2})$ and the Fourier coefficients $r_s$ of $\rho\in C^{\gamma,\alpha}$ are $O(s^{-\gamma-\alpha})$. Again, by $r_s=O(s^{-\gamma-\alpha})$, the inner sum of the third term is $O(l^{-\gamma-\alpha+1})$. Thus, for $|i-j|$ even, the third term is of order $O(n^{-1})$, if $2<\gamma+\alpha$, and of order $O(\log(n)\cdot n^{-\gamma-\alpha+1})$, if $1<\gamma+\alpha\leq 2$. To see that the last two terms are of order $O(n^{-1})$ for $|i-j|$ even, use the Cauchy-Schwarz inequality, Lemma \ref{lemma:quadrature} and the result from~\ref{app:DRB} that $\|\phi_{q,i}-\psi_{q,i}/\sqrt{n}\|_{L_2}=O(n^{-3/2})$, where $\psi_{q,i}$ is the Demmler-Reinsch basis of the Sobolev space ${\cal{W}}_\beta(M)$ given in (\ref{eq:DRSobolev}), with the tail parts $T_i(x)$ vanishing exponentially away from the boundaries. Additionally, for the last term it is used that $\left| \left(\sum_{l=1}^n  r_{|l-k|}\right )^2 \right |= O(C)$ for each $k=1,...,n$ since $\gamma+\alpha>1$. Finally, since $\rho$ is Lipschitz continuous, it follows $\rho(\pi t_{j,q})=\rho(\pi t_j)+O(n^{-1})=\rho_j+O(n^{-1})$.\\

$(ii)$ The proof is identical to $(i)$ since only the decay properties of the Fourier coefficients of $\rho$ are used. Note that $r_s=O(s^{-\beta})$ for $\rho\in\mathcal{W}_\beta(M)$.

\subsection{Simplified representation for the likelihood}\label{app:likelihood}

The log-likelihood of our model is given by
\begin{equation}
\ell_n(\lambda,q,\mathbf{R})=-\frac{n+1}{2}\log\{\mathbf{Y}^T\mathbf{R}^{-1}(\mathbf{I}_n-\mathbf{S})\mathbf{Y}+1\}+\frac{1}{2}\log|\mathbf{R}^{-1}(\mathbf{I}_n-\mathbf{S})|_+, \tag{\ref{eq:RES}}
\end{equation}
where
$$
\mathbf{R}^{-1}(\mathbf{I}_n-\mathbf{S})=\mathbf{\Phi}_q\left\{\mathbf{I}_n+\mbox{diag}(\lambda n\boldsymbol{\eta}_q)\mathbf{\Phi}_q^T\mathbf{R}\mathbf{\Phi}_q\right\}^{-1}\mbox{diag}(\lambda n\boldsymbol{\eta}_q)\mathbf{\Phi}_q^T.
$$
Suppose that Lemma~\ref{lemma:diag2} holds in such a way that we have
$$
\{\mathbf{\Phi}^T_q\mathbf{R}\mathbf{\Phi}_q\}_{ij}=\rho(\pi t_j)\delta_{ij}+\mathbb{I}\{|i-j|\mbox{ is even}\}O(n^{-1}),\;i,j=q+1,\ldots,n,
$$
which imposes some regularity constraints on the spectral density $\rho$.
In this case we can represent the matrix 
$$
\mathbf{\Phi}_q^T\mathbf{R}\mathbf{\Phi}_q=\mbox{diag}(\boldsymbol{\rho})+\mathbf{E},
$$
where $\boldsymbol{\rho}=\{\rho(\pi t_1),\ldots,\rho(\pi t_n)\}^T$ and $\mathbf{E}$ is matrix with elements 
$E_{ij}=\mathbb{I}\{|i-j|\mbox{ is even}\}O(n^{-1})$. Note that the Lemmas give the values of the matrix for $i,j=q+1,\ldots,n$, but the values for the first $q+1$ rows and columns are at most $O(1)$ (follows from the summability of the rows and columns of $\mathbf{R}$ and that $\{\mathbf{\Phi}_q\}_{ij}=O(n^{-1/2})$). \\

Now, 
\beqn
\mathbf{R}^{-1}(\mathbf{I}_n-\mathbf{S})&=&\mathbf{\Phi}_q\left\{\mathbf{I}_n+\mbox{diag}(\lambda n\boldsymbol{\eta}_q\boldsymbol{\rho})+\mbox{diag}(\lambda n\boldsymbol{\eta}_q)\mathbf{E}\right\}^{-1}\mbox{diag}(\lambda n\boldsymbol{\eta}_q)\mathbf{\Phi}_q^T\\
&=&\mathbf{\Phi}_q\left\{\mathbf{I}_n+\mbox{diag}\left(\frac{\lambda n\boldsymbol{\eta}_q}{1+\lambda n\boldsymbol{\eta}_q\boldsymbol{\rho}}\right)\mathbf{E}\right\}^{-1}\mbox{diag}\left(\frac{\lambda n\boldsymbol{\eta}_q}{1+\lambda n\boldsymbol{\eta}_q\boldsymbol{\rho}}\right)\mathbf{\Phi}_q^T.
\eeqn 
Note that the matrix
$$
\mbox{diag}\left(\frac{\lambda n\boldsymbol{\eta}_q}{1+\lambda n\boldsymbol{\eta}_q\boldsymbol{\rho}}\right)\mathbf{E}
$$
has first $q$ rows equal to zero and since the elements of the diagonal matrix are less than one, the order of this matrix is the same as that of $\mathbf{E}$. Denote 
$$
\mbox{diag}\left(\frac{\lambda n\boldsymbol{\eta}_q}{1+\lambda n\boldsymbol{\eta}_q\boldsymbol{\rho}}\right)\mathbf{E}=\frac{1}{n}\,\widetilde{\mathbf{E}},
$$
where elements of $\widetilde{\mathbf{E}}$ are of order $O(1)$ and the first $q$ rows are zero. \\
Hence,
$$
\log|\mathbf{R}^{-1}(\mathbf{I}_n-\mathbf{S})|_+=\log\left|\mbox{diag}\left(\frac{\lambda n\boldsymbol{\eta}_q}{1+\lambda n\boldsymbol{\eta}_q\boldsymbol{\rho}}\right)\right|_+-\log\left|\left(\mathbf{I}_n+n^{-1}\widetilde{\mathbf{E}}\right)D_q\right|_+,
$$
where $D_q=\mbox{diag}(0_q,1_{n-q})$. 
It is easy to see that the first term is of order $O\{n\lambda^{-1/(2q)}\}$, while the second term is $O(n^{-1})$, which follows with Hadamard's bound: $|A|^2\leq \prod_{j=1}^n\sum_{i=1}^na_{ij}^2$. Since the $ij$ elements of $(I_n+n^{-1}\widetilde{\mathbf{E}})D_q$ are of order $\delta_{ij}+O(n^{-1})$ for $i,j=q+1,\ldots,n$ and $0$ for $i,j=1,\ldots,q$
$$
\left|(\mathbf{I}_n+n^{-1}\widetilde{\mathbf{E}})D_q\right|^2_+\leq  \prod_{j=q+1}^n\{1+O(n^{-1})\}=1+O(n^{-1}),
$$
so that $\log\left|(\mathbf{I}_n+n^{-1}\widetilde{\mathbf{E}})D_q\right|^2_+=O(n^{-1})$.\\

Writing 
$$
(\mathbf{I}_n+n^{-1}\,\widetilde{\mathbf{E}})^{-1}=\mathbf{I}_n-(\mathbf{I}_n+n^{-1}\,\widetilde{\mathbf{E}})^{-1}n^{-1}\widetilde{\mathbf{E}}
$$
leads to
\begin{align*}
&\mathbf{Y}^T\mathbf{R}^{-1}(\mathbf{I}_n-\mathbf{S})\mathbf{Y}=\\&=\mathbf{B}^T\mbox{diag}\left(\frac{\lambda n\boldsymbol{\eta}_q}{1+\lambda n\boldsymbol{\eta}_q\boldsymbol{\rho}}\right)\mathbf{B}-\mathbf{B}^T(\mathbf{I}_n+n^{-1}\,\widetilde{\mathbf{E}})^{-1}n^{-1}\widetilde{\mathbf{E}}\,\mbox{diag}\left(\frac{\lambda n\boldsymbol{\eta}_q}{1+\lambda n\boldsymbol{\eta}_q\boldsymbol{\rho}}\right)\mathbf{B}.
\end{align*}
Using the definition of the matrix inverse $A^{-1}=|A|^{-1}\mbox{adj}(A)$ and Hadamard's bound as above, implies that the matrix between the two vectors in the second term has elements of order $O(n^{-1})$ for $i,j=q+1,\ldots,n$ and zero for $i,j=1,\ldots,q$. Hence,
$$
\mathbf{Y}^T\mathbf{R}^{-1}(\mathbf{I}_n-\mathbf{S})\mathbf{Y}=\sum_{i=q+1}^n\frac{B_i^2\lambda n\eta_{q,i}}{1+\lambda n \eta_{q,i}}\{1+O(n^{-1})\}-
\sum_{\substack{i,j=q+1 \\ i\neq j}}^nB_iB_jE^*_{ij}
$$
for a matrix
$$\mathbf{E}^*=(\mathbf{I}_n+n^{-1}\,\widetilde{\mathbf{E}})^{-1}(n^{-1}\widetilde{\mathbf{E}})\,\mbox{diag}\left(\frac{\lambda n\boldsymbol{\eta}_q}{1+\lambda n\boldsymbol{\eta}_q\boldsymbol{\rho}}\right),$$
which has elements of order $O(1/n)$. 
The second term in the last display is of order $O_p(1)$ by the Cauchy-Schwarz inequality.
We finally get that
\[
\mathbf{Y}^T\mathbf{R}^{-1}(\mathbf{I}_n-\mathbf{S})\mathbf{Y}=\sum_{i=q+1}^n\frac{B_i^2\lambda n\eta_{q,i}}{1+\lambda n \eta_{q,i}}\{1+O(n^{-1})\} + O_p(1).
\]


\bibliographystyle{elsarticle-num} 
\bibliography{bibSCB_additive}


%
%
%

\section*{Acknowledgements}
 The authors are greatly indebted to Dr. Christoph Lehrenfeld for helpful discussions.

\end{document}